\documentclass[10pt, conference, letterpaper]{IEEEtran}

\RequirePackage{graphics}
\usepackage{epsfig}
\usepackage{amsmath}
\usepackage{amsfonts}
\usepackage{graphicx,subfigure}
\usepackage{amssymb}
\usepackage{algorithmic}
\usepackage{algorithm,float}
\usepackage{color}
\usepackage{framed}
\usepackage{xcolor}
\usepackage{lipsum}

\newcommand{\tabincell}[2]{\begin{tabular}{@{}#1@{}}#2\end{tabular}}
\def\overbracket#1{\mathop{\vbox{\ialign{##\crcr\noalign{\kern3\p@}
\downbracketfill\crcr\noalign{\kern3\p@\nointerlineskip}
$\hfil\displaystyle{#1}\hfil$\crcr}}}\limits}
\def\underbracket#1{\mathop{\vtop{\ialign{##\crcr
$\hfil\displaystyle{#1}\hfil$\crcr\noalign{\kern3\p@\nointerlineskip}
\upbracketfill\crcr\noalign{\kern3\p@}}}}\limits}



\usepackage{cite}

\ifCLASSINFOpdf
\else
\fi

\begin{document}

\title{Challenges in Covert Wireless Communications with Active Warden on AWGN channels}


\author{\IEEEauthorblockN{
Zhihong Liu\IEEEauthorrefmark{1}, Jiajia Liu\IEEEauthorrefmark{1}, Yong Zeng\IEEEauthorrefmark{1}, Zhuo Ma\IEEEauthorrefmark{1},
Jianfeng Ma\IEEEauthorrefmark{1}, and Qiping Huang \IEEEauthorrefmark{2}}
\IEEEauthorblockA{\IEEEauthorrefmark{1}School of Cyber Engineering, Xidian University, Xi'an, China}
\IEEEauthorblockA{\IEEEauthorrefmark{2}School of Telecommunication Engineering, Xidian University, Xi'an,
China}
\IEEEauthorblockA{\IEEEauthorrefmark{5}E-mail: liujiajia@xidian.edu.cn} }


\maketitle

\begin{abstract}
Covert wireless communication or low probability of detection (LPD) communication that employs the noise or jamming signals as the cover to hide user's information can prevent a warden Willie from discovering user's transmission attempts. Previous work on this problem has typically assumed that the warden is static and has only one antenna, often neglecting an active warden who can dynamically adjust his/her location to make better statistic tests. In this paper, we analyze the effect of an active warden in covert wireless communications on AWGN channels and find that, having gathered samples at different places, the warden can easily detect Alice's transmission behavior via a trend test, and the square root law is invalid in this scenario. Furthermore, a more powerful warden with multiple antennas is harder to be deceived, and Willie's detection time can be greatly shortened.
\end{abstract}


\IEEEpeerreviewmaketitle

\section{Introduction}

Traditional network security methods based on cryptography can not solve all security and privacy problems. If a wireless node wishes to talk to others without being detected by an eavesdropper, encryption is not enough \cite{Hiding_Information}. Even a message is encrypted, the pattern of network traffic can reveal some sensitive information. If the adversary cannot ascertain Alice's transmission behavior, Alice's communication is unbreakable even if the adversary has unlimited computing and storage resources or can mount powerful quantum attacks \cite{Computer21}. On another occasion, if users hope to protect their source location privacy \cite{panda_hunter}, they also need to prevent the adversary from detecting their transmission attempts.

Covert communication, or low probability of detection (LPD) communication, has a long history.  It is always related with steganography \cite{Steganography} which hides information in covertext objects, such as images or software binary code. While steganography requires some forms of content as cover, the network covert channel requires some network protocols as its carrier \cite{covert_channel_1}. Another kind of covert communication is spread spectrum \cite{Spread_Spectrum} which is used to protect wireless communication from jamming and eavesdropping. LPD communication is also used in the underwater acoustic communication (UWAC) system for military-related applications \cite{Underwater}.

Although spread spectrum for covert communication is well-developed, its information-theoretic characteristics are unclear. Recently, another kind of physical-layer covert wireless communication that employs noise as the cover to hide user's  transmissions, is immensely intriguing to researchers. Consider a wireless communication scenario where Alice would like to talk to Bob over a wireless channel in order to not being detected by a warden Willie. Alice can use the noise in the channel instead of the statistical properties of the covert-text to hide information. Seminal work of Bash \emph{et al.} \cite{square_law} initiated the research on how the covert throughput scales with $n$, the number of channel uses in AWGN channel. It is shown that using a pre-shared key between Alice and Bob, it is possible to transmit $\mathcal{O}(\sqrt{n})$ bits reliably and covertly to Bob over $n$ channel uses such that Willie is not aware of the existence of communication.

Since covert wireless communication can provide stronger security protection, significant effort in the last few years has been devoted to achieve covertness in various network settings. If Willie has measurement uncertainty about its noise level due to the existence of SNR wall \cite{SNR}, Alice can achieve an asymptotic privacy rate which approaches a non-zero constant \cite{LDP1}\cite{Biao_He}.  In discrete memoryless channel (DMC), the privacy rate of covert communication is found to scale like the square root of the blocklength \cite{Fundamental_Limits}, and closed-form formulas for the maximum achievable covert communication rate are derived when the transmitter's channel-state information (CSI) is noncausal \cite{DMC-CSI}. To improve the performance of covert communication, Sober \emph{et al.}
\cite{jammer1} added a friendly ``jammer'' in the wireless environment to help Alice for security objectives. Soltani \emph{et al.} \cite{jammer2}\cite{DBLP:journals/corr/abs-1709-07096} considered a network scenario where there are multiple ``friendly'' nodes that can generate jamming signals to hide the transmission attempts from multiple adversaries. Liu \emph{et al.} \cite{The_Sound_and_the_Fury} and He \emph{et al.} \cite{Biao_he_TWC} studied the covert wireless communication with the consideration of interference uncertainty in a wireless network.

Previous studies on covert wireless communication are all based on the implicit assumption that the warden Willie is passive and static, which means that Willie is placed in a fixed place, judging Alice's behavior from his observations. An active Willie is a passive eavesdropper who can dynamically adjust the distance between him and Alice according to his samples. After having gathered samples at different places, Willie detect Alice's transmission attempts via a trend test. This triggers a pertinent question: "How much benefit an active Willie can gain and how do Alice deal with this situation?" Answering this question will help us better understand the benefits and limitations of covert communication in wireless networks. Besides, a more powerful Willie with multiple antennas is taken into considerations. We provide insights on the challenges in covert wireless communications with active Willie. It is harder for an active Willie to be defeated. Other secure schemes such as new artificial noise based transmission schemes, hybrid beamforming techniques need to be developed to secure wireless transmission \cite{Safeguarding_5G}.

\section{Background and Related Work}
\subsection{Covert Throughput and Pre-shared Secret}
Bash, Goeckel, and Towsley's work \cite{square_law} is the first work that puts information theoretic bound on covert wireless communication. A square root law is found over noisy AWGN channels: Alice can only transmit $\mathcal{O}(\sqrt{n})$ bits reliably and covertly to Bob over $n$ uses of channels. The reason for the sub-linearly covert throughput is that Alice can only conceal transmissions in the standard deviation of the channel noise. However, Bash's scheme needs a large pre-shared secret $\mathcal{O}(\sqrt{n}\log n)$ bits in $n$ channel uses. In a different model, if Alice transmits only once in a long sequence of possible transmission slots and Willie does not know the time of transmission attempts, Alice can reliably transmit $\mathcal{O}(\min\{\sqrt{n\log(T(n))}, n\})$ bits to Bob with a slotted AWGN channel \cite{time1}.

To eliminate the need for a long shared secret between Alice and Bob, Che \emph{et al.} \cite{CheBJ13}\cite{CheBCJ14} studied covert communication over a binary symmetric channel (BSC). There is no pre-shared secret hidden from Willie. The only asymmetry between Bob and Willie is that Willie's channel is worse than Bob's, and the best privacy rate Alice can obtain is a constant rate. In a discrete memoryless multiple-access channel, Arumugam and Bloch \cite{Keyless} showed that, if the channel to Bob is better than the one to Willie, Alice can covertly communicate to Bob on the order of $\sqrt{n}$ bits per $n$ channel uses without using a pre-secret. Another method which can achieve a nonzero privacy rate is discussed in \cite{LDP1}.  Leveraging results on the phenomenon of SNR wall, Lee \emph{et al.} found that a nonzero privacy rate is also possible and the pre-shared secret is not needed. Furthermore, the recent work of Arrazola \emph{et al.}\cite{PhysRevA.97.022325} demonstrates that the amount of key consumed in a covert communication protocol may be smaller than the transmitted key, thus leading to secure secret key expansion.

\subsection{Channel Models}
The first studied channel model is AWGN channel, the standard model for a free-space RF channel, where the signal is corrupted by the addition of a sequence of i.i.d. zero-mean Gaussian random variables.  The square root law was found over AWGN channels \cite{square_law} and slow fading channels \cite{8406961}. Yan \emph{et al.} \cite{Delay-Intolerant} first studied delay-intolerant covert communications in AWGN channels with a finite block length. They found that $n = N$ is optimal to maximize the amount of information bits that can be transmitted covertly, where $n$ is the actual number of channel uses and $N$ is the maximum allowable number of channel uses.

Recently, covert communication has been extended to various channel models. In \cite{LDP1}, Lee \emph{et al.} extended their work from AWGN Rayleigh SISO channel to MIMO channels with infinite samples when an
eavesdropper employs a radiometer detector and has uncertainty about his noise variance. In discrete memoryless channels, Wang \emph{et al.} \cite{Fundamental_Limits} found that the privacy rate of covert
communication scales like the square root of the blocklength. Arumugam and Bloch \cite{Keyless} studied covert communication in a discrete memoryless multiple-access channel, and extended their work to a $K$-user multiple access channel \cite{K-User} in which $K$ transmitters attempt to communicate covert messages reliably to a legitimate receiver. Che \emph{et al.} \cite{CheBJ13}\cite{CheBCJ14} first considered covert communication over a binary symmetric channel. Besides, Bash \emph{et al.} \cite{Quantum} studied covert communication in quantum channels, and even generalized the results with similar throughput scaling. Soltani \emph{et al.} \cite{renewal} studied the covert communications on renewal packet channels where the packet timings of legitimate users are governed by a Poisson point process.

\subsection{Codes for Covert Communications}
The classical coder for covert communications in AWGN channel is random coding. Alice takes input in blocks of size $M$ bits and encodes them into codewords of length $n$. She independently generates $2^{nR}$ codewords and constructs a codebook which is used as the secret key shared between Alice and Bob. In practice, Bash \cite{Bash_PHD} proposed to use any error-correction codes to reliably transmit $\mathcal{O}(\sqrt{n})$ covert bits using  $\mathcal{O}(\sqrt{n}\log n)$ pre-shared secret bits. However, it is still challenging to share such a long key in advance. A more practical method is using a short secret key as the initial key for a stream-cipher, such as Trivium \cite{Preneel2005Trivium}, to generate a long key.

For covert communications over asynchronous discrete memoryless channels, Freche \emph{et al.} \cite{Polar_Codes1}\cite{Polar_Codes} proposed a binary polar code scheme which can achieve good performance close to the random coding scheme with lower complexity. Bloch \cite{Bloch} discussed covert communications from a resolvability perspective, and developed an alternative coding scheme to achieve the covertness. In \cite{Computationally_Efficient}, Zhang \emph{et al.} designed computationally efficient codes with provable guarantees on both reliability and covertness over BSCs which can achieve the best known throughput.

\subsection{Multi-hop Covert Communication and Shadow Network}
Previous work on covert communication mainly focus on the  performance analysis of 1-hop systems, while the performance analysis on multi-hop systems remains largely unknown.

In \cite{NaNA}, Wu \emph{et al.} considered covert communication in a two-hop wireless system where Alice
communicates with Bob via a relay.  Their results indicate that LPD communication can be guaranteed if the maximum throughput is limited to $\mathcal{O}(\sqrt{n})$ bits in $n$ channel uses. In \cite{Greedy_Relay}, Hu \emph{et al.} studied the possibility and achievable performance of covert communication in one-way relay networks with a greedy relay. In their setting, the relay is greedy and opportunistically transmits its own information covertly, while Alice tries to detect this covert transmission.

Sheikholeslami \emph{et al.} \cite{8315147} considered multi-hop covert communication over a moderate size
network and in the presence of multiple collaborating Willies. They developed efficient algorithms to find optimal paths with maximum throughput and minimum delay. With the aid of friendly jammers, Soltani \emph{et al.} \cite{jammer2}\cite{DBLP:journals/corr/abs-1709-07096} studied a network scenario where there are multiple ``friendly'' nodes that can generate artificial noise to impair wardens' ability to detect transmissions.

The classical covert wireless communication hides the signal in the noise. Although the ambient noise is unpredictable to some extent, the aggregated interference in a wireless network is more difficult to be predicted. Shabsigh \emph{et al.} \cite{MILCOM_2016} used stochastic geometry to study the design of Ad-Hoc covert networks that can hide their transmissions in the spectrum of primary networks. He \emph{et al.} \cite{Biao_he_TWC} studied covert communication in wireless networks in which Bob and Willie are subject to uncertain shot noise from interferers. Liu \emph{et al.} \cite{The_Sound_and_the_Fury} also considered the covert communication in a noisy wireless network. Their results show that Alice can reliably and covertly transmit $\mathcal{O}(\log\sqrt{n})$ bits in $n$ channel uses if the distance between Alice and Willie is larger than a bound which is only related to $n$. From the network perspective, the communications can be hidden in the noisy wireless networks, and what Willie sees is merely a shadow wireless network.

We would like to point out that all the results discussed above are based on the implicit assumption that Willie is passive, static and has only one antenna, while in this paper we seek to understand whether a more powerful Willie indeed affects the covert wireless communications and the challenges Alice will be confronted with.


\section{System Model}\label{ch_2}
\subsection{Channel Model}
Consider a wireless communication scene where Alice (A) wishes to transmit messages to Bob (B) covertly. A warden Willie (W) is eavesdropping over the wireless channel and trying to find whether or not Alice is transmitting. We adopt the wireless channel model similar to \cite{square_law}. Each node, legitimate node or eavesdropper, is equipped with a single omnidirectional antenna (Willie with multiple antennas will be discussed in section \ref{ch_multi_ant}). All wireless channels are assumed to suffer from discrete-time AWGN with real-valued symbols, and the wireless channel is modeled by large-scale fading with path loss exponent $\alpha$.

Let the transmit power employed for Alice be $P_0$, and $s^{(A)}$ be the real-valued symbol Alice transmitted which is a Gaussian random variable $\mathcal{N}(0,1)$ by employing a Gaussian codebook. Suppose $z^{(B)}\sim \mathcal{N}(0, \sigma^2_{B,0})$ is the AWGN at Bob, and $z^{(W)}$ is the AWGN at Willie with $z^{(W)}\sim \mathcal{N}(0, \sigma^2_{W,0})$. Assume Bob and Willie experience the noise with the same power, i.e., $\sigma^2_{B,0} = \sigma^2_{W,0}=\sigma^2_0$. Then, the signal seen by Bob and Willie when Alice is transmitting can be represented as follows,
\begin{eqnarray}
  y^{(B)}  &\equiv & \sqrt{\frac{P_0}{d_{A,B}^{\alpha}}} \cdot s^{(A)}+z^{(B)}~~\sim~ \mathcal{N}(0,\sigma^2_{B}) \label{eq_1}\\
  y^{(W)}  &\equiv & \sqrt{\frac{P_0}{d_{A,W}^{\alpha}}} \cdot s^{(A)}+z^{(W)}~~\sim~ \mathcal{N}(0,\sigma^2_{W}) \label{eq_2}
\end{eqnarray}
and
\begin{equation}\label{eq_4_4}
    \sigma^2_{B}=\frac{P_0}{d_{A,B}^{\alpha}}+\sigma^2_0, ~~ \sigma^2_{W}=\frac{P_0}{d_{A,W}^{\alpha}}+\sigma^2_0
\end{equation}
where $d_{A,B}$ and $d_{A,W}$ are the Euclidean distances between Alice and Bob, Alice and Willie,
respectively.

\subsection{Covert Communication}
To transmit a message to Bob covertly and reliably, Alice can use the classical encoder in \cite{square_law} and suppose that Alice and Bob have a shared secret of sufficient length, based on which Alice selects a codebook from an ensemble of codebooks. As to Willie, without knowing the secret key, he cannot decide with arbitrarily low probability of detection error that whether his observation is a signal transmitted by Alice or the noise of the channel.

The codebook Alice chooses is low power codebook, and any error-correction code can be used to construct a covert communication system \cite{Bash_PHD}. In this paper, we assume that Alice and Bob randomly select the symbol periods that they will use for their transmission by flipping a biased coin $n$ times, with probability of heads $r=\mathcal{O}(\frac{1}{\sqrt{n}})$.  On average, $\eta=\mathcal{O}(\sqrt{n})$ symbol periods is selected. Bob simply ignores the discarded symbol periods, however, Willie cannot do so and thus observes mostly noise. Furthermore, Alice randomly generate an $\eta$-symbol vector $\mathbf{k}$ secretly from Willie, and XOR the encoded message ($\eta$ symbols) with this secret vector $\mathbf{k}$. Then Alice transmits on $\eta$ symbol periods selected. XORing by vector $\mathbf{k}$ prevents Willie's exploitation of the error correction code's structure to detect Alice (rather than protects the message content).

\subsection{Active Willie}
In \cite{square_law} and \cite{DBLP:journals/corr/abs-1709-07096}, Willie is assumed to be passive and static, which means that Willie is placed in a fixed place, eavesdropping and judging Alice's behavior
from his $n$ channel samples $y^{(W)}_1, y^{(W)}_2, \cdots, y^{(W)}_n$ with each sample $y^{(W)}_i\sim
\mathcal{N}(0,\sigma^2_{W})$. Based on the sampling values, Willie employs a radiometer as his detector, and decides whether Alice is transmitting or not.

\begin{figure}
\centering \epsfig{file=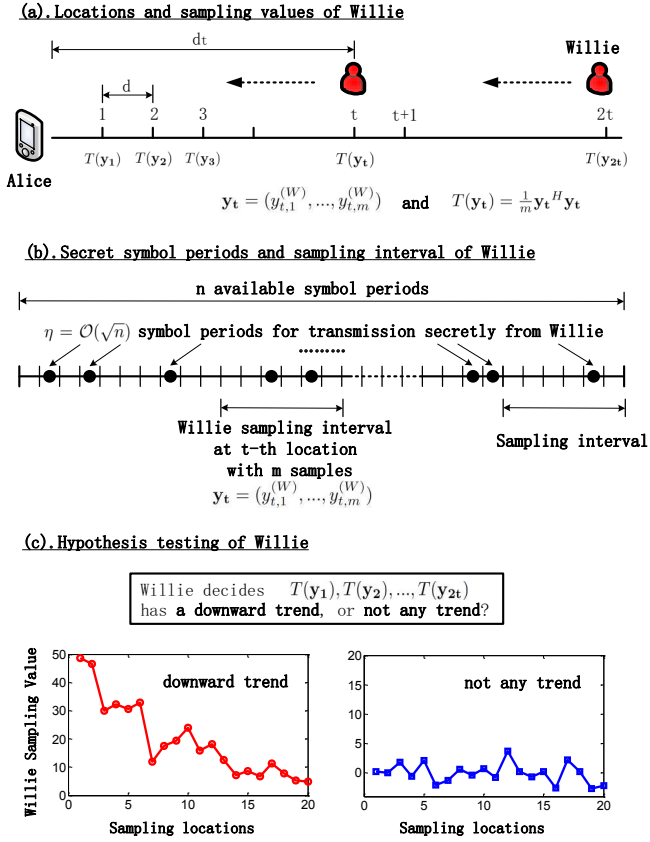, height=4.4in}
\caption{Covert wireless communication in the presence of an active Willie who leverages a trend analysis to detect Alice's transmission attempts.}\label{framework}
\end{figure}

The system framework with an active Willie is depicted in Fig. \ref{framework}.  Willie detects Alice's behavior at $2t$ different locations (each location is $d$ meters apart). At each location he gathers $m$ samples. For example, at $t$-th location (with the distance $d_t$ between Alice and Willie), Willie's samples can be presented as a vector
\begin{equation}
    \mathbf{y_t}=(y_{t,1}^{(W)}, y_{t,2}^{(W)}, \cdots, y_{t,m}^{(W)})
\end{equation}
where each sample $y^{(W)}_{t,m}\sim \mathcal{N}(0,\sigma^2_{W_t})$, and $\sigma^2_{W_t}=P_t
+\sigma^2_0=\frac{P_0}{{d_t}^{\alpha}}+\sigma^2_0$.

The average sampling value at $t$-th location can be calculated as follows
\begin{equation}
    T(\mathbf{y_t})=\frac{1}{m}\mathbf{y_t}^H\mathbf{y_t} =\frac{1}{m}\sum^m_{k=1}y_{t,k}^{(W)}*y_{t,k}^{(W)}.
\end{equation}
Therefore Willie will have a sampling value vector $\mathbf{T}$, consisting of $2t$ values at different locations,
\begin{equation}
    \mathbf{T}=(T(\mathbf{y_1}), T(\mathbf{y_2}), \cdots,  T(\mathbf{y_{2t}}))
\end{equation}
Then Willie decides whether $\mathbf{T}$ has a downward trend or not. If the trend analysis shows a
downward trend for given significance level $\beta$, Willie can ascertain that Alice is transmitting  with probability $1-\beta$.

\subsection{Hypothesis Testing}
To find whether Alice is transmitting or not, Willie has to distinguish between the following two hypotheses,
\begin{eqnarray}
\mathbf{H_0}&:& \text{there is not any trend in vector $\mathbf{T}$;} \\
                      & & y^{(W)}  \equiv  z^{(W)}  \nonumber\\
\mathbf{H_1}&:& \text{there is a downward trend in vector $\mathbf{T}$.}\\
                      & & y^{(W)}  \equiv  \sqrt{\frac{P_0}{d_{A,W}^\alpha}}\cdot s^{(A)}+ z^{(W)} \nonumber
\end{eqnarray}

Given the sampling value vector $\mathbf{T}$, Willie can leverage the Cox-Stuart test \cite{cox} to detect the
presence of trend. The idea of the Cox-Stuart test is based on the comparison of the first and the second half
of the samples. If there is a downward trend, the observations in the second half of the samples should be
smaller than in the first half.  If there is not any trend, Willie should expect only small differences between the first and the second half of the samples due to randomness. The differences of samples can be calculated as follows
\begin{eqnarray}
  \Delta_1 &=& T(\mathbf{y_1})-T(\mathbf{y_{t+1}}) \nonumber\\
  \Delta_2 &=& T(\mathbf{y_2})-T(\mathbf{y_{t+2}}) \nonumber\\
     & &  \cdot\cdot\cdot\cdot\cdot\cdot \nonumber\\
  \Delta_t &=& T(\mathbf{y_t})-T(\mathbf{y_{2t}}) \nonumber
\end{eqnarray}
Let $sgn(\Delta_i)=1$ for $\Delta_i<0$, $sgn(\Delta_i)=0$ for $\Delta_i\geq 0$. The test statistic of the
Cox-Stuart test on the vector $\mathbf{T}$ is
\begin{equation}
\mathbf{T}_{\Delta<0}=\sum^t_{i=1}sgn(\Delta_i)
\end{equation}

\begin{figure}
\centering \epsfig{file=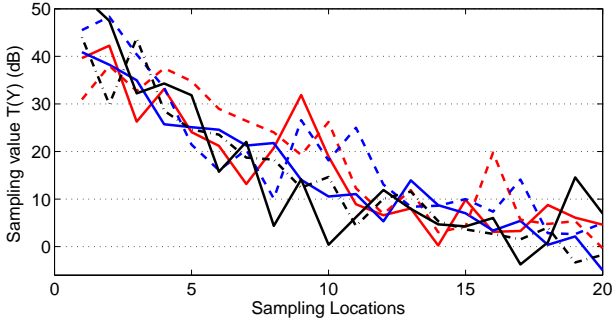, height=1.7in}
\caption{The sampling values at different locations when Alice is transmitting with the transmission probability $r=0.5$. Here a bounded path loss law is used, $l(x)=\frac{1}{1+\parallel x\parallel^\alpha}$. The transmission power $P_0$ of Alice is 30dB, links experience unit mean Rayleigh fading, and $\alpha=3$. The spacing between sampling locations $d=0.2$m. }\label{trend01}
\end{figure}

Given a significance level $\beta$ and the binomial distribution $\mathbf{b}\sim b(t,0.5)$, Willie rejects the null hypothesis $\mathbf{H_0}$ and accepts the alternative hypothesis $\mathbf{H_1}$ if $\mathbf{T}_{\Delta<0}<\mathbf{b}(\beta)$ which means a downward trend is found with probability larger than $1-\beta$, where $\mathbf{b}(\beta)$ is the  quantile of the binomial distribution $\mathbf{b}$. According to the central limit theorem, if $t$ is large enough ($t>20$), an approximation $\mathbf{b}(\beta)=1/2[t+\sqrt{t}\cdot\Phi^{-1}(\beta)]$ can be applied, where $\Phi^{-1}(\beta)$ is the
$\beta$-quantile function of the standard normal distribution. Therefore, if
\begin{equation}\label{qq}
    \mathbf{T}_{\Delta<0}<\frac{1}{2}[t+\sqrt{t}\cdot\Phi^{-1}(\beta)]
\end{equation}
Willie can ascertain that Alice is transmitting with probability larger than $1-\beta$ for the significance level $\beta$ of test. Fig. \ref{trend01} shows examples of the sampling values at different locations when Alice is transmitting. The downward trend of the signal power is obvious when Alice is transmitting with certain transmission probability.

The parameters and notation used in this paper are illustrated in Table \ref{tab_1}.

\begin{table}
\caption{parameters and notation}
\begin{tabular}{|l|l|}
  \hline
  $\mathbf{Symbol}$ & $\mathbf{Meaning}$ \\
  \hline
  $P_0$ & Transmission power of Alice \\
  \hline
  $t$ & \tabincell{l}{Number of differences in Cox-Stuart test \\ with $2t$ sampling values.}  \\
  \hline
  $r$ & Alice's transmit probability  \\
  \hline
  $\alpha$ & Path loss exponent \\
  \hline
  $m$ & Number of samples in a sampling location \\
  \hline
  $P_i$ & Willie's received power at $i$-th sampling location \\
  \hline
  $d_i$ & Distance between Alice and Willie's $i$-th location \\
    \hline
  $d$ & Spacing between sampling points \\
    \hline
  $s^{(A)}$ & Alice's signal  \\
  \hline
  $y^{(B)}$, $y^{(W)}$ & Signals Bob and Willie observe  \\
  \hline
  $z^{(B)}$, $z^{(W)}$ & (Bob's, Willie's) background noise \\
  \hline
  $\mathbf{y_t}$ & Willie's samples at $t$-th location \\
  \hline
  $T(\mathbf{y_t})$ & Willies's sampling value at $t$-th location \\
  \hline
  $\mathbf{T}$ & Willie's sampling value vector \\
  \hline
  $\lambda$ & Density of the network \\
  \hline
  $\mathcal{N}(\mu,\sigma^2)$ & \tabincell{l}{Gaussian distribution with mean $\mu$ and  variance $\sigma^2$} \\
  \hline
  $\mathbf{T}_{\Delta<0}$ &  Test statistic of the Cox-Stuart test \\
  \hline
  $\beta$ &  Significance level of testing \\
  \hline
  $\mathbf{E}[X]$ & Mean of random variable $X$ \\
   \hline
   $\mathbf{Var}[X]$ & Variance of random variable $X$ \\
   \hline
   $\mathbf{cov}[X,Y]$ & Covariance of random variable $X$ and $Y$ \\
   \hline
  $\Phi^{-1}(\beta)$ & $\beta$-quantile function of $\mathcal{N}(0, 1)$ \\
   \hline
   $I_i$ & Interference signal at $i$-th sampling location  \\
  \hline
  $\sigma^2_{I_i}$ & Power of interference signal at $i$-th sampling location  \\
  \hline
  $\rho_{I_i,I_j}$ & Correlation of interference signals $I_i$ and $I_j$  \\
  \hline
  $\rho_{\sigma^2_{I_i},\sigma^2_{I_j}}$ & Correlation of $\sigma^2_{I_i}$ and $\sigma^2_{I_j}$  \\
  \hline
\end{tabular}\label{tab_1}
\end{table}

\section{Active Willie Attack} \label{ch_3}
This section discusses the covert wireless communication in the presence of an active Willie. As illustrated in Fig. \ref{framework},  Willie detects Alice's behavior  at $2t$ different locations. At each location he gathers $m$ samples and then calculates the sampling value at this location. At $i$-th location (with the distance $d_i$ between Alice and Willie), Willie's samples are a vector
\begin{equation}
    \mathbf{y_i}=(y_{i,1}^{(W)}, y_{i,2}^{(W)}, \cdots, y_{i,m}^{(W)})
\end{equation}
with
\begin{equation}
y^{(W)}_{i,m} =
\begin{cases}
\mathcal{N}(0,\sigma^2_{0})& \mathbf{X}=0 \\
\mathcal{N}(0,\sigma^2_{W_i})& \mathbf{X}=1
\end{cases}
\end{equation}
where $\mathbf{X}$ is a random variable, $\mathbf{X}=1$ if Alice is transmitting in the current symbol period, $\mathbf{X}=0$ if Alice is silent, and the transmission probability $\mathbb{P}\{\mathbf{X}=1\}=r$. $\sigma^2_{W_i}=P_i +\sigma^2_0=\frac{P_0}{{d_i}^{\alpha}}+\sigma^2_0$.

Then the sampling value at this location is
\begin{equation}
    T(\mathbf{y_i})=\frac{1}{m}\mathbf{y_i}^H\mathbf{y_i}=\frac{1}{m}\sum_{k=1}^m [y_{i,k}^{(W)}]^2
\end{equation}
and
\begin{equation}
[y_{i,k}^{(W)}]^2 =
\begin{cases}
\sigma^2_0\chi^2(1)& \mathbf{X}=0 \\
\sigma^2_{W_i}\chi^2(1)& \mathbf{X}=1
\end{cases}
\end{equation}
where $\chi^2(1)$ is the chi-squared distribution with 1 degree of freedom.



In this paper, we focus on the circumstance of $n\rightarrow\infty$, which allows Willie to observe a large number of samples at each location, i.e., $m$ is large enough.  Because $\{[y_{i,1}^{(W)}]^2, [y_{i,2}^{(W)}]^2, \cdots, [y_{i,m}^{(W)}]^2\}$ is a sequence of independent random variables, and satisfies Lindeberg's condition (the proof is placed in Appendix A), based on the Lindeberg central limit theorem, we have
\begin{equation}
\sum_{k=1}^m [y_{i,k}^{(W)}]^2\sim \mathcal{N}\biggl(\sum_{k=1}^m \mathbf{E}[[y_{i,k}^{(W)}]^2],\sum_{k=1}^m\mathbf{Var}[[y_{i,k}^{(W)}]^2]\biggr)
\end{equation}
with
\begin{eqnarray}
\mu_{i}&=&\mathbf{E}[[y_{i,k}^{(W)}]^2] = r\sigma^2_{W_i} + (1-r)\sigma^2_{0} \\
\sigma^2_{i}&=&\mathbf{Var}[[y_{i,k}^{(W)}]^2] = 3[r\sigma^4_{W_i}+(1-r)\sigma^4_{0}] \\
                                & & -[r\sigma^2_{W_i}+(1-r)\sigma^2_{0}]^2 \nonumber
\end{eqnarray}

Then the sampling value at this location is
\begin{equation}
    T(\mathbf{y_i})=\frac{1}{m}\sum_{k=1}^m [y_{i,k}^{(W)}]^2\sim\mathcal{N}\biggl(\mu_{i},\frac{\sigma^2_{i}}{m}\biggr)
\end{equation}

As to Willie, his received signal strength at $i$-th location is $P_i=P_0d_i^{-\alpha}=P_0(i\cdot d)^{-\alpha}$ which is a decreasing function of the distance $d_i$. At first Willie monitors the environment, if he detects the anomaly with $P_{i}\geq\sigma_0^2$, Willie then approaches Alice to carry out more stringent testing. According to the setting, we have
\begin{equation}
    P_1>P_2>\cdots>P_{2t-1}>P_{2t}=\sigma_0^2
\end{equation}

With $2t$ sampling values $T(\mathbf{y_t})$  at different locations, Willie can decide whether $(T(\mathbf{y_1}), T(\mathbf{y_2}), \cdots, T(\mathbf{y_{2t}}))$ has a downward trend or not via the Cox-Stuart test. The differences satisfy the following distribution
\begin{equation}
\Delta_i=T(\mathbf{y_i})-T(\mathbf{y_{t+i}})\sim\mathcal{N}(\mu_{\Delta_i},\sigma^2_{\Delta_i})
\end{equation}
where
\begin{equation}\label{mu_delta}
\mu_{\Delta_i}=\mu_{i}-\mu_{t+i}=\sigma_0^2A(i,t,r,\alpha)
\end{equation}
\begin{equation}\label{sigma_delta}
\sigma^2_{\Delta_i} =\frac{\sigma^2_{i}+\sigma^2_{t+i}}{m} = \frac{\sigma_0^4}{m}B(i,t,r,\alpha)
\end{equation}
and
\begin{equation}\label{mu_delta_1}
A(i,t,r,\alpha)=r\biggl[\biggl(\frac{2t}{i}\biggr)^\alpha -  \biggl(\frac{2t}{t+i}\biggr)^\alpha  \biggr]
\end{equation}
\begin{eqnarray}\label{sigma_delta_1}
B(i,t,r,\alpha) &=& 3\biggl[ r\biggl(\biggl( \frac{2t}{i}\biggr)^\alpha+1\biggr)^2 + (1-r) \biggr] - \nonumber\\
                               &  & - \biggl[ r\biggl(\frac{2t}{i}\biggr)^\alpha+1\biggr]^2 \nonumber\\
                               &  & + 3\biggl[ r\biggl(\biggl( \frac{2t}{t+i}\biggr)^\alpha+1\biggr)^2 + (1-r) \biggr] - \nonumber\\
                               &  & - \biggl[ r\biggl(\frac{2t}{t+i}\biggr)^\alpha+1\biggr]^2
\end{eqnarray}

The probability that the difference $\Delta_i=T(\mathbf{y_i})-T(\mathbf{y_{t+i}})<0$ ($1\leq i\leq t $) can be estimated as follows,
\begin{equation}\label{ss27}
   \mathbb{P}\{\Delta_i<0\} = \mathbb{P}\{T(\mathbf{y_i})<T(\mathbf{y_{t+i}})\}=\Phi\biggl(-\frac{\mu_{\Delta_i}}{\sigma_{\Delta_i}}\biggr)
\end{equation}

However, the number of negative  differences in $\Delta_1, \Delta_2, ..., \Delta_t$ is $t$ Poisson trials where the success probabilities $\mathbb{P}\{\Delta_i<0\}$ ($i = 1,2,\cdots, t$) differ among the trials, it has no standard distribution. We find that $\frac{\mu_{\Delta_i}}{\sigma_{\Delta_i}}$ decreases with $i$. Therefore the number of negative differences in $\Delta_1, \Delta_2, ..., \Delta_t$ can be upper bounded as follows
\begin{equation}\label{df}
    \mathbf{T}_{\Delta<0}=\sum^t_{i=1}\mathbb{P}\{\Delta_i<0\}\leq t\mathbb{P}\{\Delta_t<0\}=t\Phi\biggl(-\frac{\mu_{\Delta_t}}{\sigma_{\Delta_t}}\biggr)
\end{equation}
and
\begin{equation}
\frac{\mu_{\Delta_t}}{\sigma_{\Delta_t}}=\sqrt{m}\cdot\kappa(r,\alpha)
\end{equation}
where
\begin{equation}
\kappa(r,\alpha)\approx\frac{2^\alpha-1}{2}\cdot r
\end{equation}
when $r$ is small enough.

If $t$ is large enough ($t>20$), the following inequality holds
\begin{equation}\label{qaqa}
    \mathbf{T}_{\Delta<0}\leq t\Phi(-\sqrt{m}\cdot\kappa(r,\alpha))< \frac{1}{2}[t+\sqrt{t}\cdot\Phi^{-1}(\beta)].
\end{equation}

Therefore, given any small significance level $\beta>0$, if the number of locations $t$ satisfies
\begin{equation}
  t>\biggl[\frac{\Phi^{-1}(\beta)}{1-2\Phi(-\sqrt{m}\cdot\kappa(r,\alpha))}\biggr]^2,
\end{equation}
Willie can distinguish between two hypotheses $\mathbf{H_0}$ and  $\mathbf{H_1}$ with probability $1-\beta$.

According to Equ. (\ref{qaqa}), Willie with certain $m$ and $t$ can detect Alice's transmission if the transmission probability of Alice $r$ is larger than the threshold as follows
\begin{equation}\label{threshold}
    r\geq r_{max}=\frac{2}{2^\alpha-1}\cdot\frac{1}{\sqrt{m}}\cdot\Phi^{-1}\biggl[ 1-\frac{1}{2}\biggl( 1+\frac{\Phi^{-1}(\beta)}{\sqrt{t}}\biggr) \biggl]
\end{equation}

\begin{figure}
\centering \epsfig{file=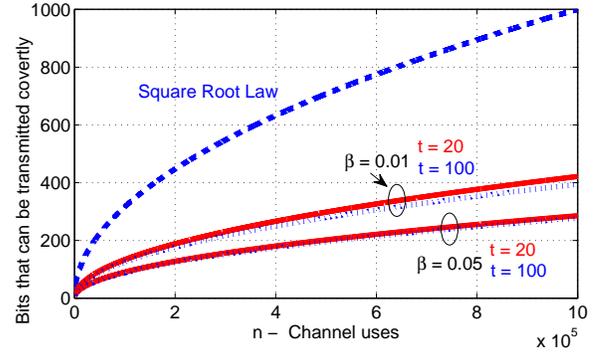, height=1.9in}
\caption{Number of bits that can be transmitted covertly. Here $n = t*m$ and $\alpha=4$.}\label{n_bit}
\end{figure}

This may be a pessimistic result since it demonstrates that for given certain $m$ (or $t$) and $r$, Willie can find certain $t$ (or $m$) to detect Alice's transmission behavior. If Alice uses the channel a finite number of times, $n$, and Willie samples the channel $m*t=n$ times, Alice can only transmit $\mathcal{O}(n*r_{max})$ bits covertly. Fig.\ref{n_bit} shows the number of bits that can be transmitted covertly versus the number of channel uses, $n$, with different system configurations. We can observe that the number of covert bits increases with channel uses and is much less than the square root law. We also observe that for fixed values of $n$ and $t$, by increasing the parameter $\beta$, the number of covert bits will be decreased and the effect of $t$ on the number of covert bits for given $\beta$ is very small.

\subsection{Willie's detection strategy}
As discussed above, given certain sampling locations $t$ and the number of samples $m$ at each location, Willie can ascertained with significant high probability that Alice is transmitting or not. However, if Willie chooses large $m$ and $t$ to detect Alice's transmission behavior when Alice's transmission probability $r$ is small, the sample collection procedure will last long time. Furthermore, without any knowledge of Alice, Willie does not know how to select the optimal parameter $m$ and $t$. Small parameters will make Willie missing detection when Alice's transmission probability is very small, large parameters, however, will make the detection time unaccepted.

\textbf{Detection Strategy}: To make detection efficiently and quickly, Willie divides the detection procedure into multiple rounds. At each round, Willie samples the channel at each detection location with a small $m$. At the end of each round, Willie tests Alice's behavior with the samples gathered in this round and all other rounds before. If a downward tendency is found in the sampling values, Willie declares that Alice is transmitting. Otherwise, Willie carries out the next round of sampling and testing until a conspicuous behavior is found.

In this way, Willie can quickly find Alice's transmission when the transmission probability is not small. In the case that Alice intends to transmit a short message, if Willie does not adopt the round based detection strategy and chooses a large parameter $m$, it will take Willie a very long time to find this transmission. What is worse, the successful detection probability will decrease since more noise is involved into the sampling values. In practice, because Willie has no information about Alice, it is necessary to set a maximum round value, $round_{max}$. If Willie does not find any suspicious behavior of Alice through $round_{max}$ rounds of inspection, he then restarts the detection from scratch. The $round_{max}$ should be set properly, smaller value will lead to miss detection of Alice's transmission with very small transmission probability, larger value on the other hand will involve in more channel noise if Alice's transmission message is very short.

\subsection{Successful Detection Probability}
Next we estimate the successful detection probability of Willie. Suppose Willie samples Alice's transmission signal at $2t$ locations, each with $m$ samples at a round. With $2t$ sampling values $T(\mathbf{y_t})$  at different locations, Willie then try to decide whether $(T(\mathbf{y_1}), T(\mathbf{y_2}), \cdots, T(\mathbf{y_{2t}}))$ has a downward trend or not via the Cox-Stuart test. He calculates the differences
$\Delta_i=T(\mathbf{y_i})-T(\mathbf{y_{t+i}})$ ($i = 1..t$), and constructs a  test statistic
$\mathbf{T}_{\Delta<0}=\sum^t_{i=1}sgn(\Delta_i)$ (where $sgn(\Delta_i)=1$ for $\Delta_i<0$,
$sgn(\Delta_i)=0$ for $\Delta_i\geq 0$). Given a significance level $\beta$ and the binomial distribution
$\mathbf{b}\sim b(t,0.5)$, he rejects the null hypothesis $\mathbf{H_0}$ and accept the alternative hypothesis $\mathbf{H_1}$ if $\mathbf{T}_{\Delta<0}<\mathbf{b}(\beta)$.

According to Equ. (\ref{ss27}), we have
\begin{equation}\label{sas}
   \mathbb{P}\{\Delta_i<0\} = \Phi\biggl(-\frac{\mu_{\Delta_i}}{\sigma_{\Delta_i}}\biggr)
\end{equation}
and
\begin{equation}\label{sas}
   \mathbb{P}\{\Delta_1<0\}<\cdots <\mathbb{P}\{\Delta_t<0\}
\end{equation}
Becasue $\mathbf{T}_{\Delta<0}$ is the sum of $t$ Poisson trials with different success probabilities. Let $\mu=\textbf{E}(\mathbf{T}_{\Delta<0})$ denote the sum's expected value. According to multiplicative Chernoff bound, for any $\delta>0$,
\begin{equation}\label{Chernoff}
   \mathbb{P}\{\mathbf{T}_{\Delta<0}<(1+\delta)\mu\}\geq 1-\biggl[ \frac{e^\delta}{(1+\delta)^{(1+\delta)}} \biggr]^\mu
\end{equation}

For given significance $\beta$, let $\mathbf{b}(\beta)=(1+\delta)\mu$, then the probability that Willie can detect Alice's transmission behavior, $\mathbf{P}_{succ}$, can be lower bounded as follows
\begin{eqnarray}
  \mathbf{P}_{succ} &=& \mathbb{P}\{\mathbf{T}_{\Delta<0}<\mathbf{b}(\beta)\}  \nonumber \\
     &\geq & 1-\biggl[ \frac{e^\delta}{(1+\delta)^{(1+\delta)}} \biggr]^\mu
    \label{q2}
\end{eqnarray}
for any $\delta = \mathbf{b}(\beta)/\mu-1>0$, where $\mathbf{b}(\beta)=1/2[t+\sqrt{t}\cdot\Phi^{-1}(\beta)]$, and
\begin{equation}
    \mu=\sum_{i=1}^t\mathbb{P}\{\Delta_i<0\}=\sum_{i=1}^t\Phi\biggl(-\frac{\mu_{\Delta_i}}{\sigma_{\Delta_i}}\biggr)
\end{equation}
where $\mu_{\Delta_i}$ and $\sigma_{\Delta_i}$ are defined in (\ref{mu_delta}) and (\ref{sigma_delta}), and
\begin{equation}
    \frac{\mu_{\Delta_i}}{\sigma_{\Delta_i}}=\frac{A(i,t,r,\alpha)\sqrt{m}}{\sqrt{B(i,t,r,\alpha)}}\sqrt{R}
\end{equation}
Here $A(i,t,r,\alpha)$ and $B(i,t,r,\alpha)$ are defined in (\ref{mu_delta_1}) and (\ref{sigma_delta_1}), and $R$ represents the number of rounds, $m$ is the number of samples per round at a location.

\begin{figure} \centering
\subfigure[]{ \epsfig{file=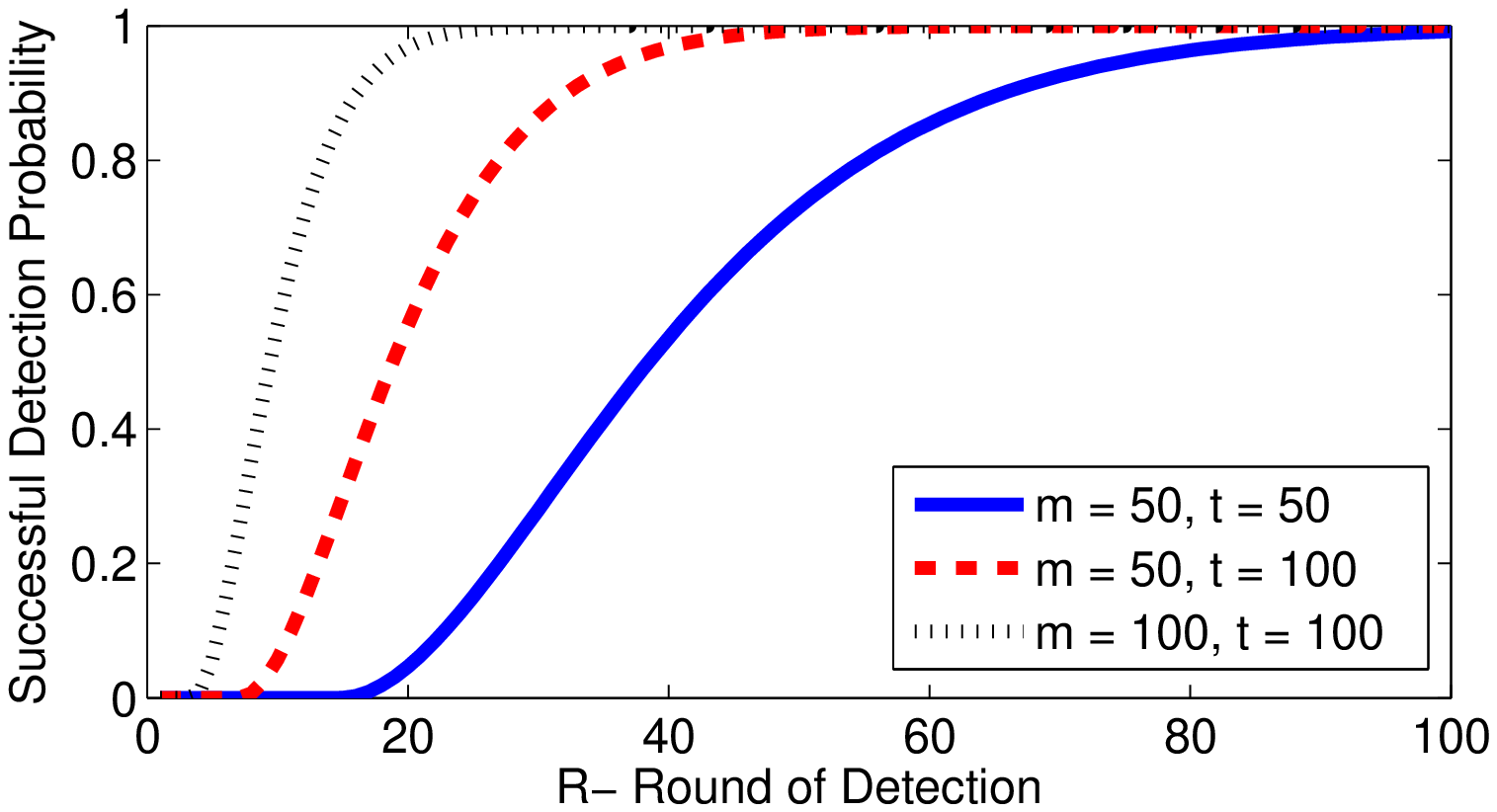, height=1.5in }}
\subfigure[]{ \epsfig{file=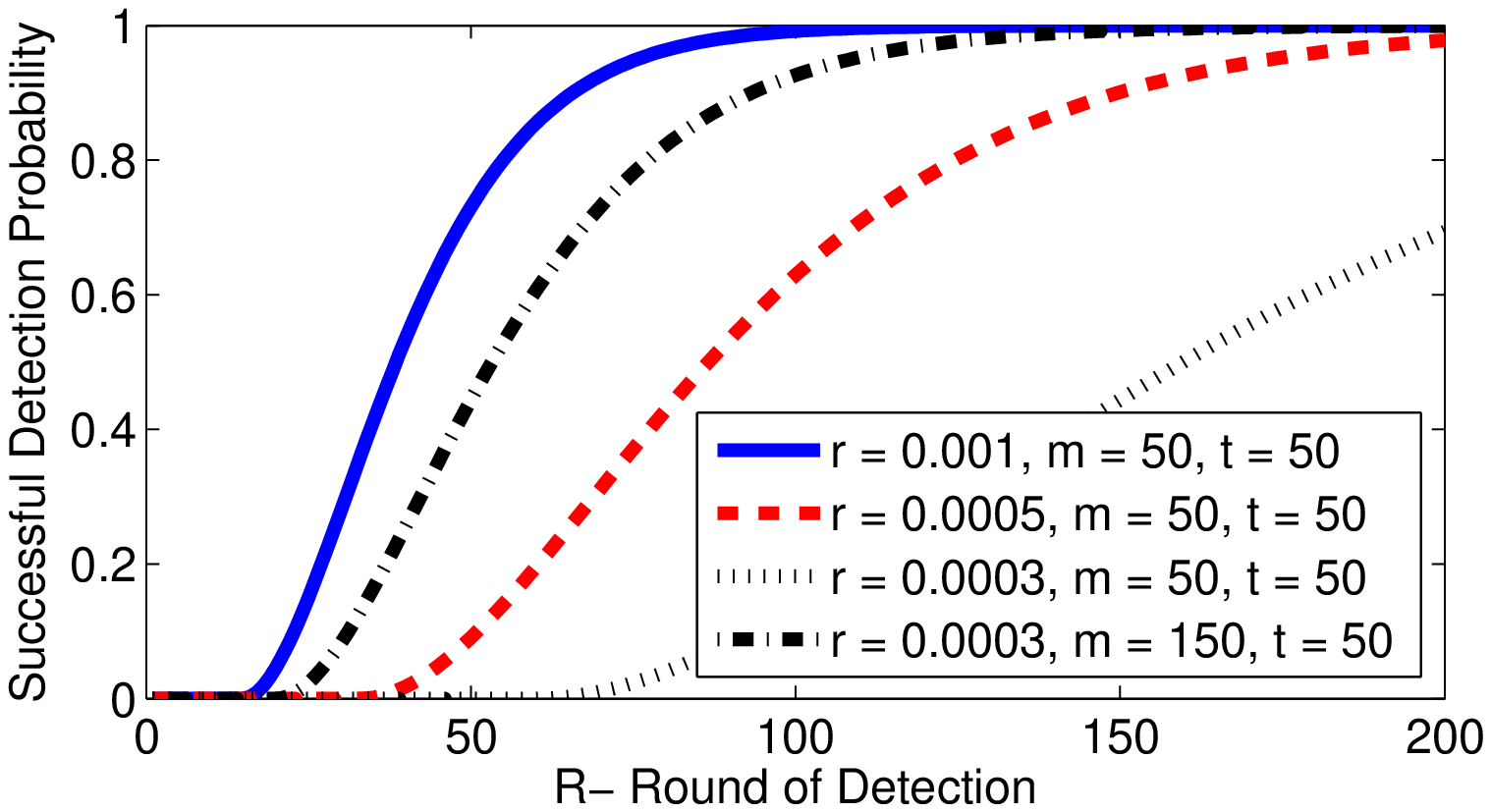, height=1.5in} }
\caption{The successful detection probability of Willie versus rounds of detection for
(a) different parameters $m$ and $t$ when $r = 0.001$, and (b) different transmission probability $r$. Here $\alpha=4$ and $\beta=0.01$. } \label{succ_d_p}
\end{figure}

Fig. \ref{succ_d_p} shows the successful detection probability versus rounds of detection for different parameters. What can be clearly seen in the figures is the rapid increase in the successful detection probability when more detection rounds are taken. For given transmission probability of Alice, Fig. \ref{succ_d_p}(a) illustrates that more detection locations or more samples at each location will have higher detection probability. If Alice decreases her transmission probability, her transmission behavior is harder to be found, since Willie's successful detection probability decreases rapidly with $r$, as depicted in Fig. \ref{succ_d_p}(b).

However, no matter how low Alice's transmission probability is, Willie can ascertain Alice's transmitting attempt with probability $1-\beta$ for any small $\beta$ and certain number of rounds. This may be a pessimistic result since it demonstrates that Alice cannot resist the attack of active Willie and the square root law does not hold in this situation.

\section{Countermeasures to Active Willie} \label{ch_4}
If Alice has knowledge about Willie, such as Willie's location, she can decrease her transmission power when Willie is approaching. However,  Alice may be a small and simple IoT device who is not able to perceive the environmental information.

\begin{figure} \centering
\subfigure[]{ \epsfig{file=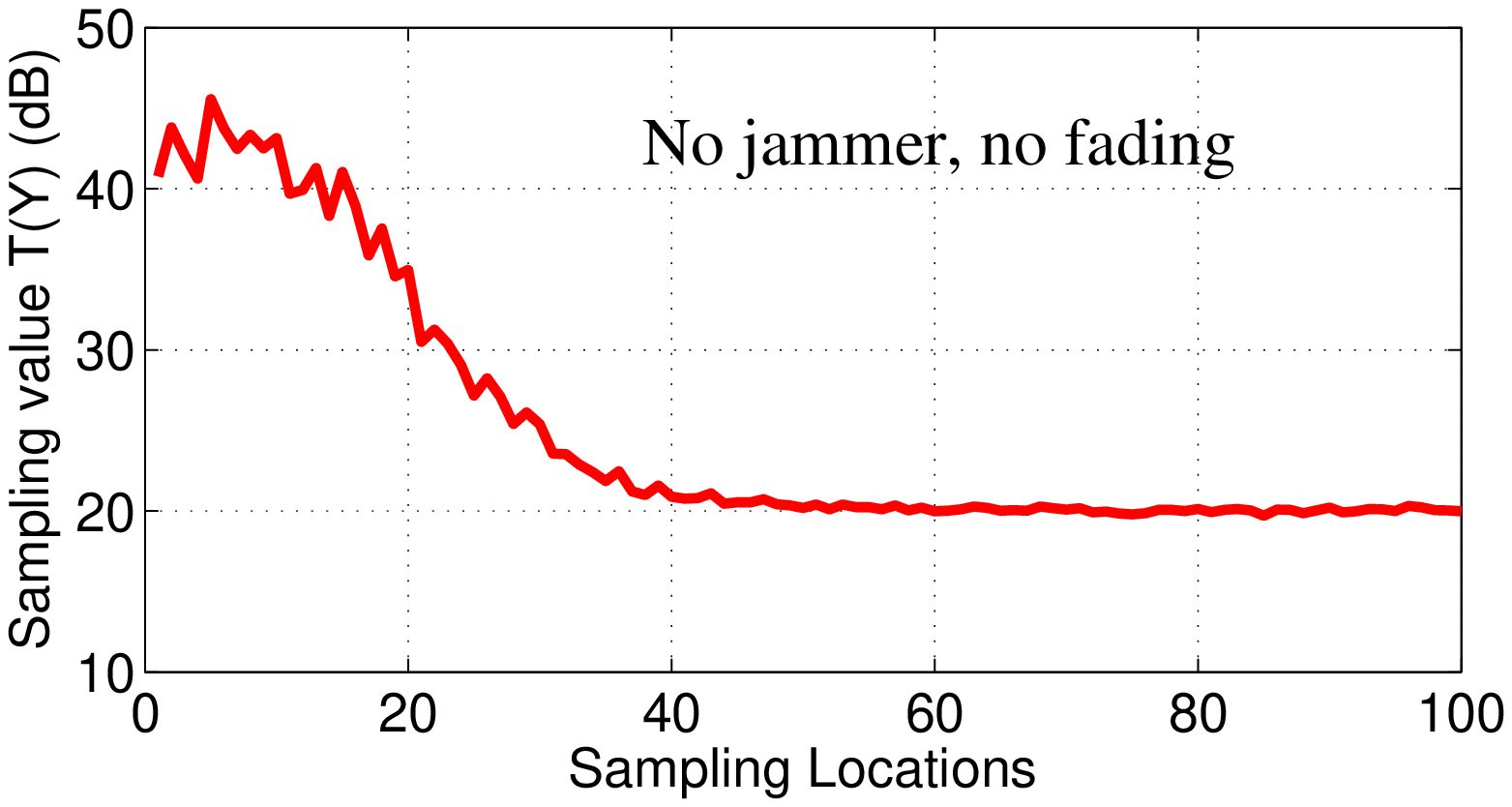, height=0.8in }}
\subfigure[]{ \epsfig{file=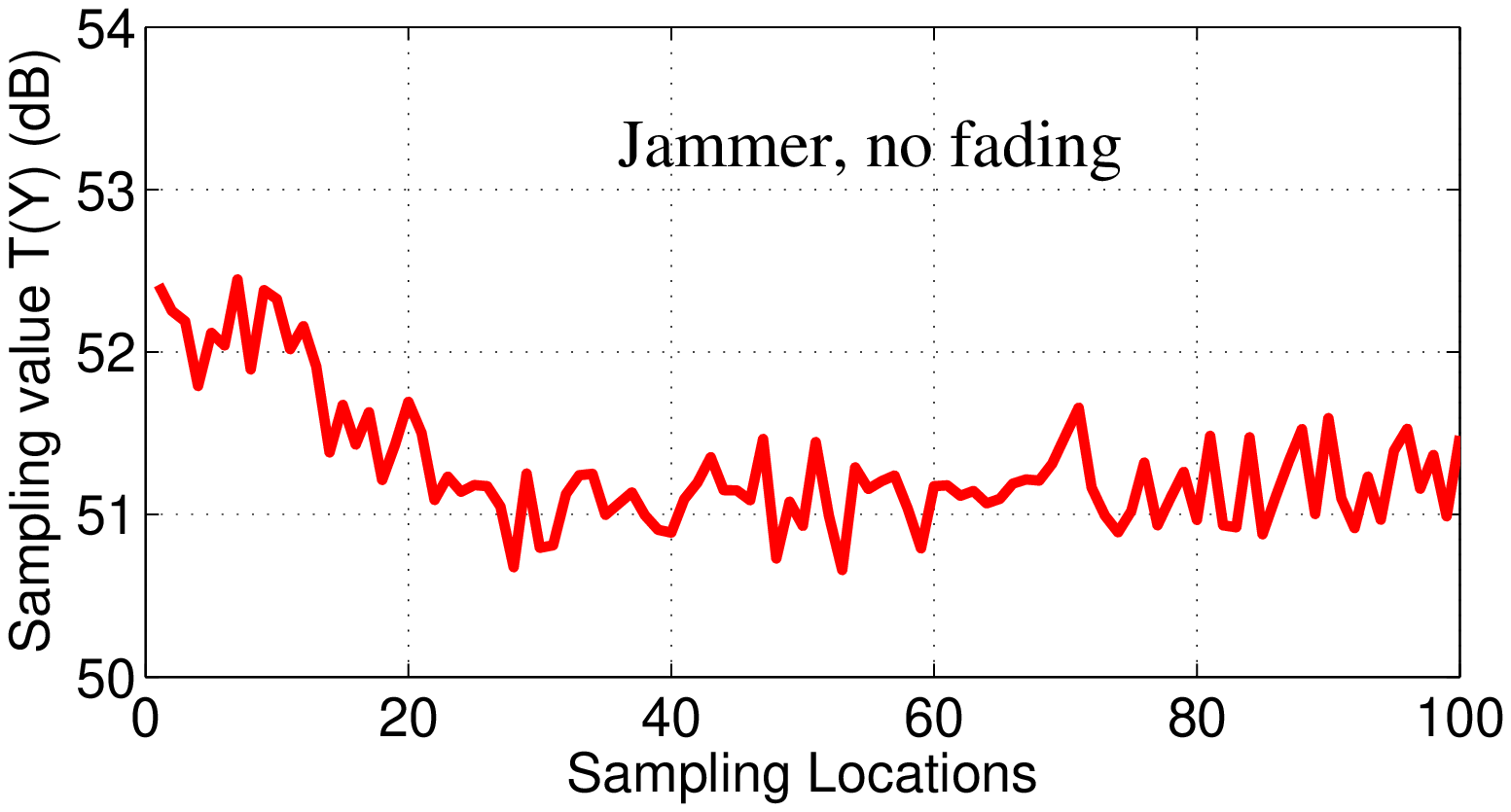, height=0.8in} }
\subfigure[]{ \epsfig{file=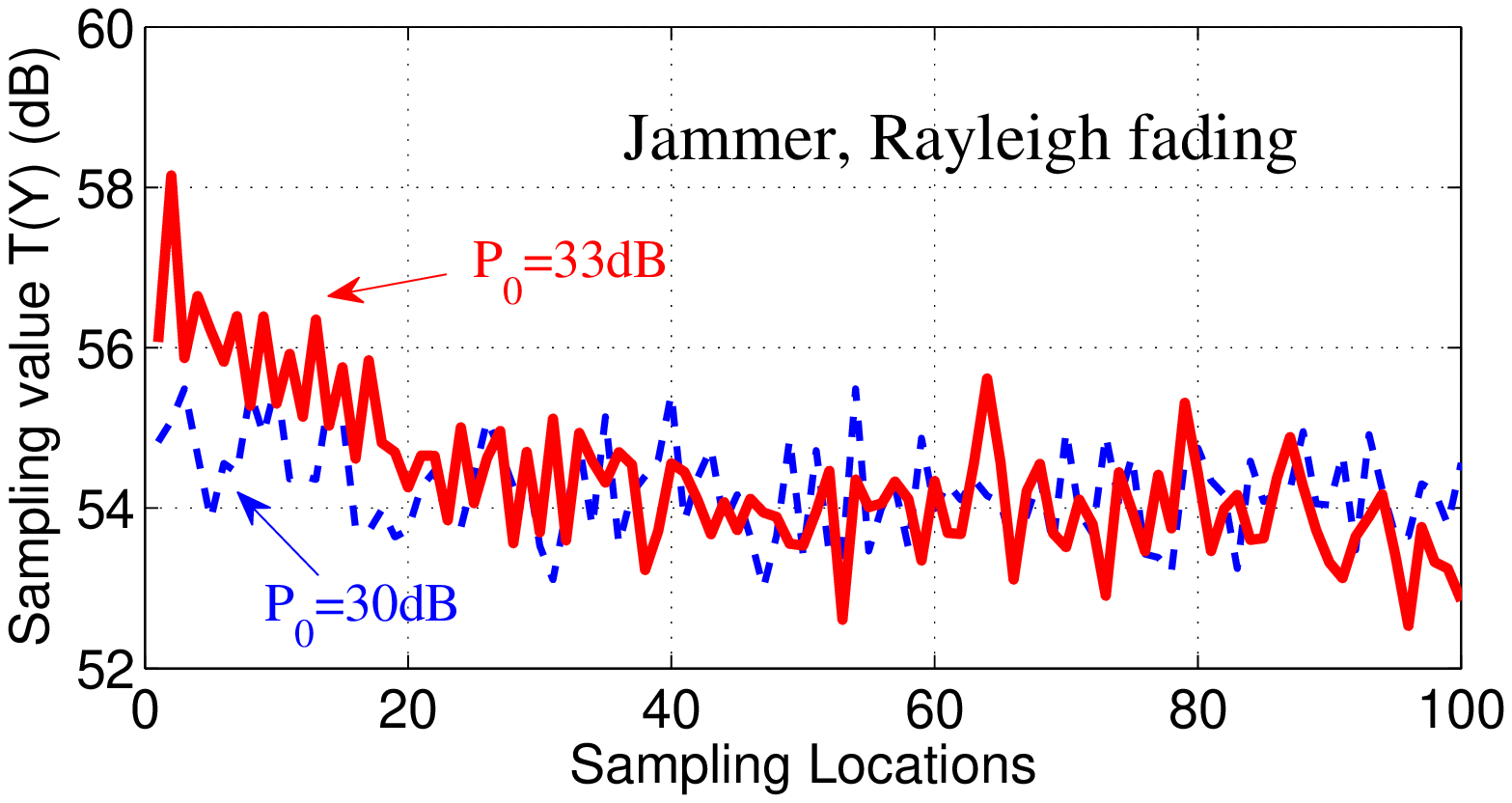, height=0.8in} }
\subfigure[]{ \epsfig{file=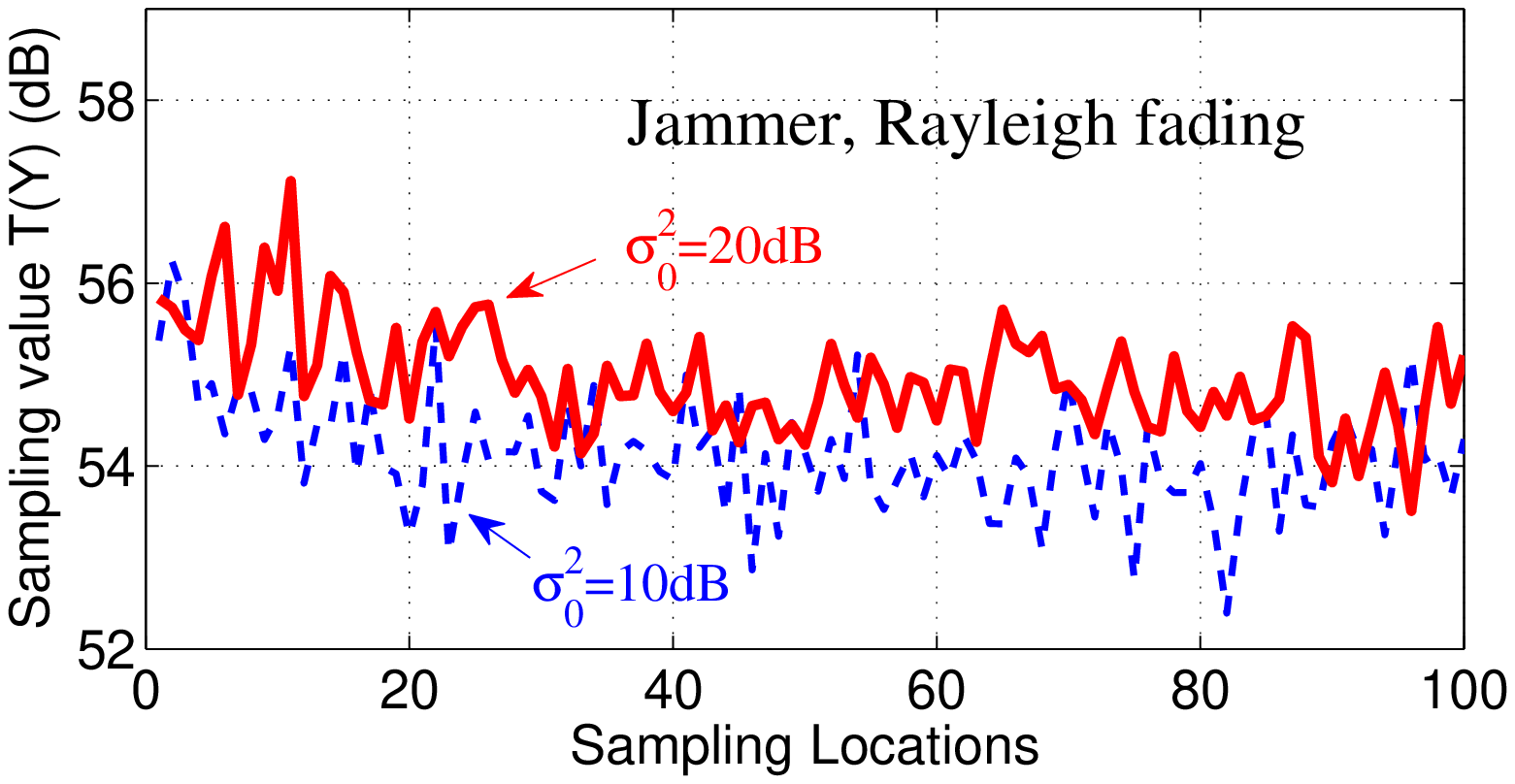, height=0.8in }}
\subfigure[]{ \epsfig{file=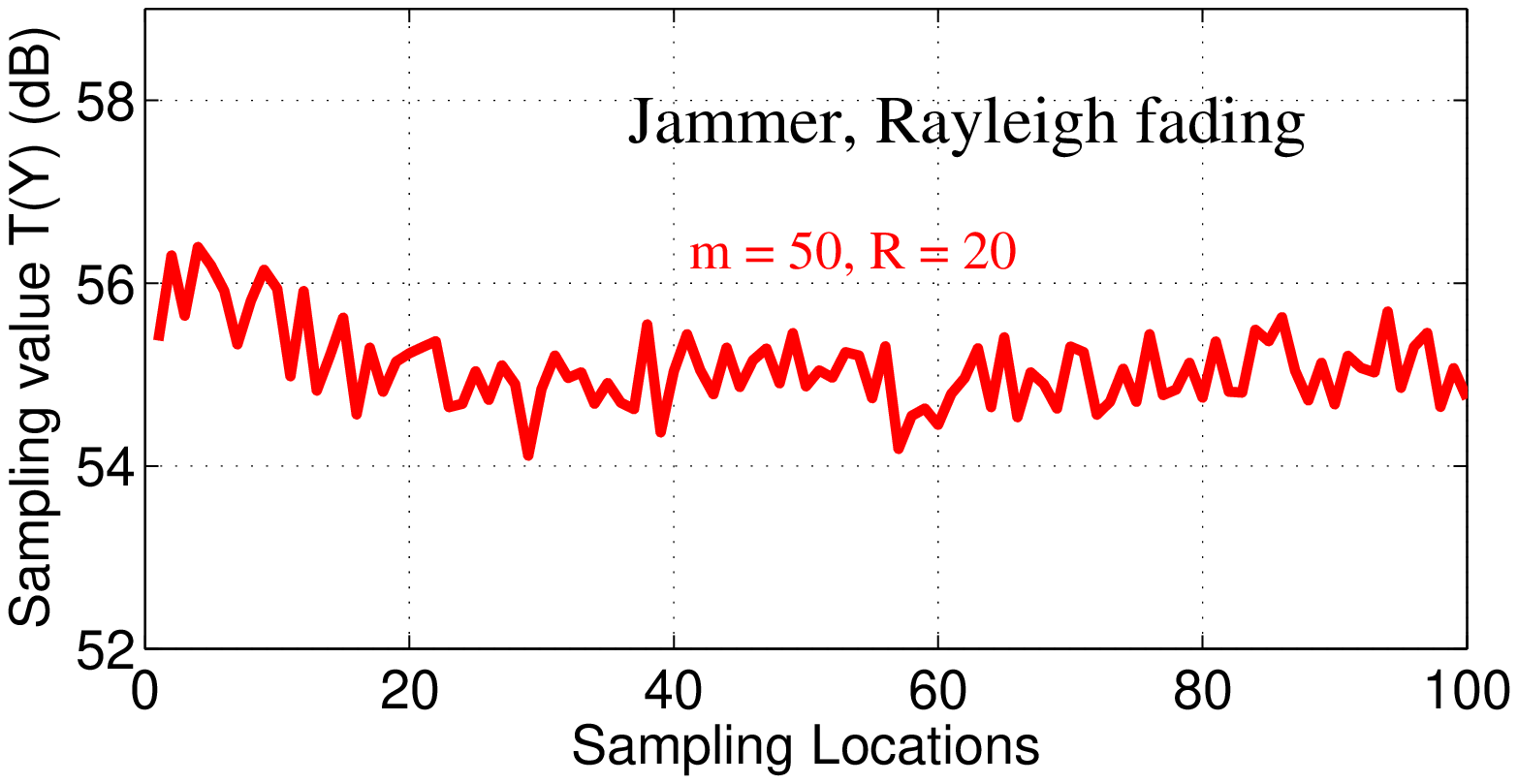, height=0.8in} }
\subfigure[]{ \epsfig{file=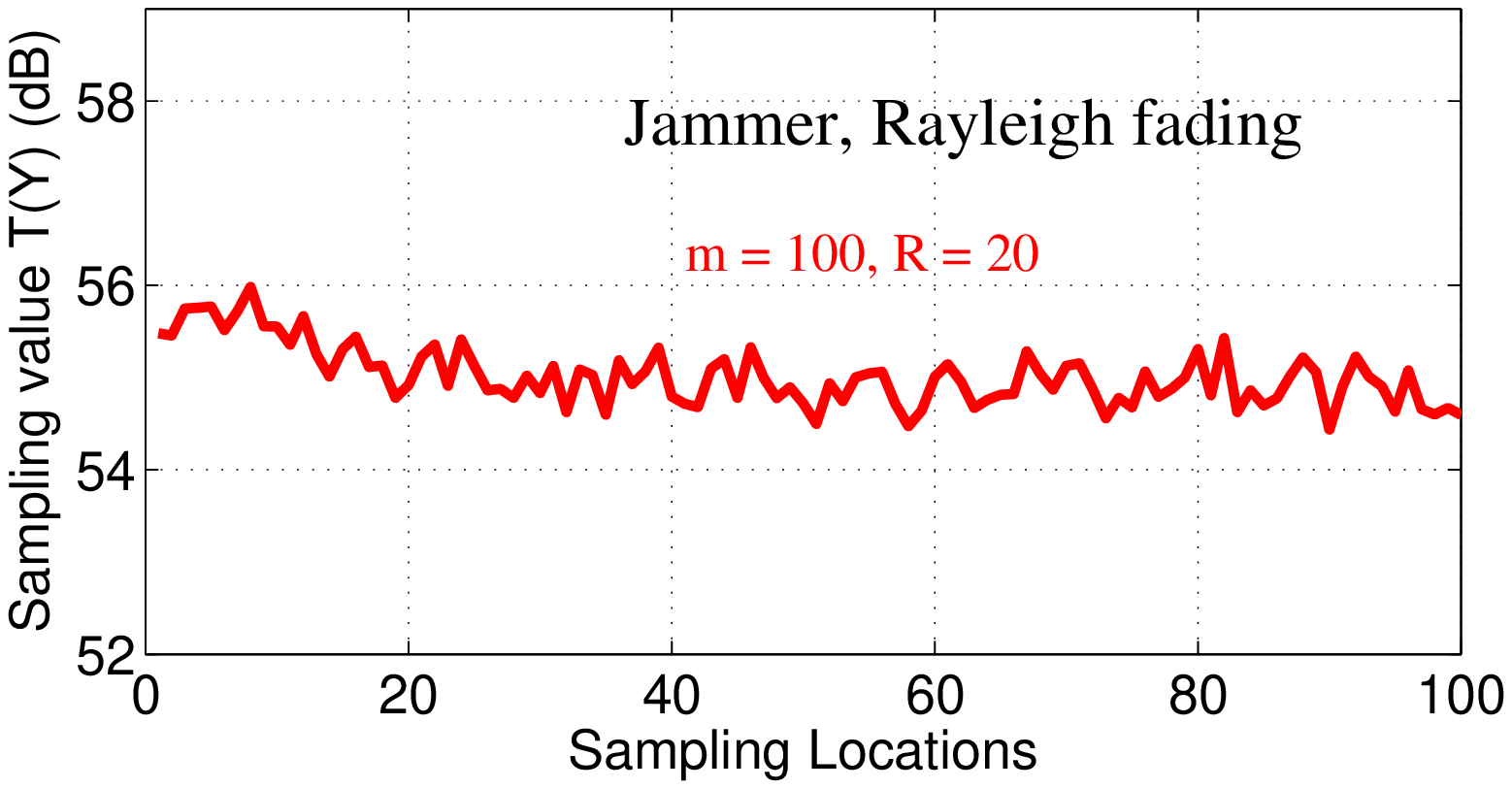, height=0.8in} }
\caption{The sampling values at different locations with the transmit probability $r=0.02$, and $d = 10m$. The path loss model is $l(x)=\frac{1}{1+\parallel x\parallel^\alpha}$ with $\alpha=4$.
(a) No jammer and no Rayleigh fading in channels are taken into considerations. The transmit power of Alice is $P_0=30$dB, the number of samples at each location per round is $m = 50$, the noise power is $\sigma^2_0=10$dB, the sampling values are computed through $R=20$ rounds of detection.
(b) A friendly jammer is introduced into the network to emit interference signal to help Alice, no Rayleigh fading in channel is considered. The jammer moves with Willie and keeps the distance between him and Willie $d_j,w = 1$m, his jamming signal power is set to $P_j=30$dB.
(c) A friendly jammer and the unit mean Rayleigh fading are taken into considerations. Alice uses different transmit power to communicate with Bob. Red line represents $P_0=33$dB, the dotted blue line denotes $P_0=30$dB.
(d) A friendly jammer and the unit mean Rayleigh fading are taken into considerations. Different noise power levels are compared. Red line represents the noise power $\sigma^2_0=20$dB, the dotted blue line denotes $\sigma^2_0=10$dB.
(e) The sampling values at different locations when $m=50$ and $R=20$;
(f) The sampling values at different locations when $m=100$ and $R=20$.
}
\label{measures}
\end{figure}

In practice, to confuse Willie, Alice should decrease her transmission power. Clearly, when Alice's transmission power $p_0$ decreases, Willie's uncertainty increases. However, if Willie can move close to Alice, the downward tendency of the signal power is still exist, although this tendency becomes weakening. Fig.\ref{measures}(a) depicts this tendency of the sampling values at different locations. Another intuitive countermeasure is to further confuse Willie with the aid of a friendly jammer, as discussed in \cite{jammer1}. Although a jammer can help Alice  for security objectives by artificially increasing the noise level of Willie, it cannot change the tendency of the sampling values. As illustrated in Fig.\ref{measures}, if a jammer and Alice do not coordinate and the jammer randomly changes his/her power of the Gaussian noise in each slot of signals, the magnitude of signal fluctuation increases. However, there is still a downtrend, even in a fading channel. If Willie can collect more samples at each location, he can obtain a smoother sampling values.

\subsection{Interference of Static Network}
For a static and passive Willie, to discriminate the actual transmitter from the other in a network is a difficult task, provided that there is no obvious radio fingerprinting of transmitters can be exploited \cite{radio_fingerprinting}. For the reasons discussed above, Alice cannot defeat the attack of an active Willie, even an uninformed jammer is introduced in the environment. Next we investigate the covert performance if Alice is put into a dense static wireless network.

\begin{figure} \centering
\subfigure[Static wireless network topology for simulation.]{ \epsfig{file=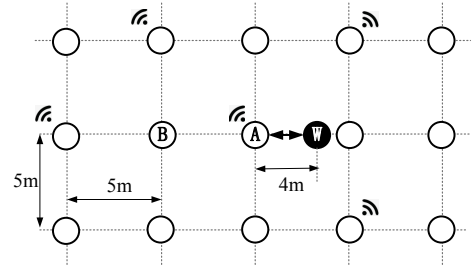, height=1.4in }}
\subfigure[]{ \epsfig{file=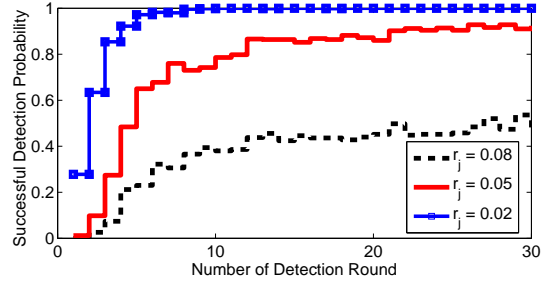, height=1.5in} }
\caption{(a) The static wireless network is deployed in a $400\times 400 m^2$ space; (b) The probability of Willie's successful detection versus the number of detection rounds for different jamming probability. Here the results are averaged over 500 experiments, and at each round, parameters $m=50$, $r = 0.02$, $\alpha=4$ and $\beta=0.05$, $d = 4m$, $t = 30$, $\sigma^2=10$dB, $p_0 = p_j = 30$dB. } \label{Static}
\end{figure}

Given a static wireless network deployment as depicted in Fig.\ref{Static}(a), wireless nodes are deployed in a grid network with 5 meters apart. Each node transmits signal randomly and independently with the same transmit power. As shown in Fig.\ref{Static}(b), with the increase of jamming probability $r_j$, Willie's successful detection probability decreases. This makes sense because Willie will receive more interference if the surrounding nodes transmits with higher probability. However, as more rounds of detection are taken, Willie's successful detection probability increases, implying that the interference in a static network is not enough to hide Alice's transmission behavior.

\subsection{Interference of Mobile Network}
The above discussion shows that the interference in a static network cannot completely confuse Willie. Willie could approach Alice as close as possible, and ensure that there is no other node located closer to Alice than him. Next we put an active Willie in a mobile network to test whether or not Willie will be bewildered by the interference from a large number of mobile nodes.

\begin{figure}
\centering \epsfig{file=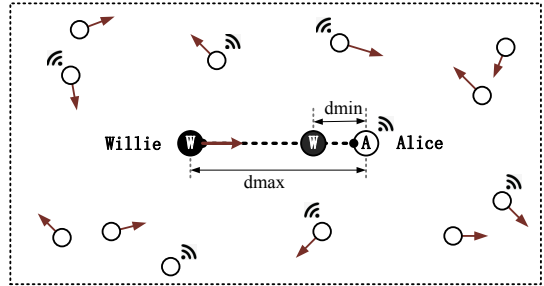, height=1.5in} \caption{Mobile network topology for simulation.}\label{mobile_deployment}
\end{figure}

We consider a mobile network with 800 nodes deployed in a $400\times 400 m^2$ space. As depicted in Fig.\ref{mobile_deployment}, Alice is placed in the center of this region, Willie is $dmax$ meters away from Alice. However Willie cannot get too close to Alice, $dmin$ is the minimum distance between them. In a wireless network, some wireless nodes are probably placed on towers, trees, or buildings, Willie cannot get close enough as he wishes. Each mobile node transmits signal randomly and independently with the jamming probability $r_j$. For the movement of nodes, we adopt the Random Walk Mobility Model \cite{mobile_model} used for simulation of mobile ad-hoc network. The Random Walk Mobility Model mimics the unpredictable movements of many objects in nature. In our simulation, mobile nodes are randomly deployed in the area. Each node randomly selects a direction and move at a random speed uniformly selected from [0, 6]m/s. The new direction and speed are randomly selected every second.

\begin{figure} \centering
\centering \epsfig{file=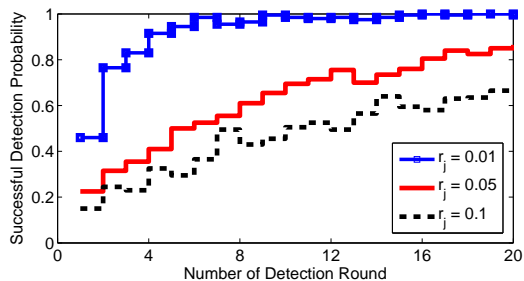, height=1.5in}
\caption{The successful detection probability of Willie versus the number of rounds of detection for different jamming probability $r_j$. Here the results are averaged over 200 experiments, and at each round, parameters $t = 30$, $m=50$, and $r = 0.01$. The distance between Alice and Willie is $dmin = 0.5$m and $dmax = 4$m, Alice's transmit power and the jamming power of each mobile node is $p_0 = p_j = 30$dB.} \label{mobile1}
\end{figure}

Fig.\ref{mobile1} shows Willie's successful detection probability versus the rounds of detection for different jamming probability $r_j$. As more detection rounds are taken, Willie's successful detection probability increases as well. The higher the jamming probability $r_j$ mobile nodes adopt, the lower the detection probability since more interference is involved in the sampling values.

\begin{figure} \centering
\subfigure[]{ \epsfig{file=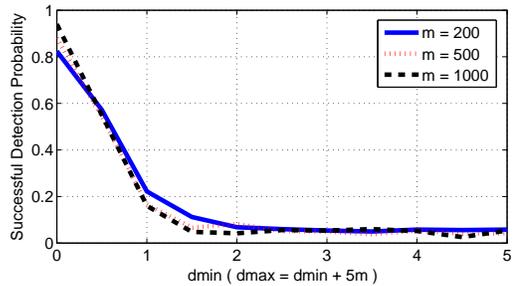, height=1.5in }}
\subfigure[]{ \epsfig{file=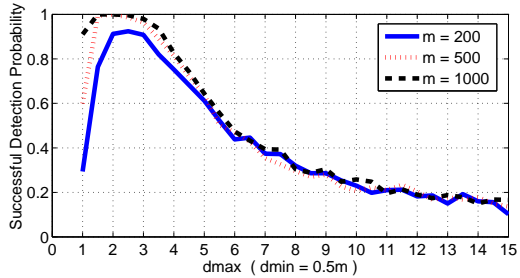, height=1.5in} }
\caption{The successful detection probability of Willie versus $dmin$ and $dmax$ for different $m$. Here the results are averaged over 200 experiments, and at each round, parameters $r = r_j = 0.01$, $\alpha=4$ and $\beta=0.05$. (a) The minimum distance $dmin$ increases, and $dmax = dmin + 5$m; (b) The minimum $dmin$ is set to be 0.5m, $dmax$ grows.} \label{mobile23}
\end{figure}

Fig.\ref{mobile23}(a) and Fig.\ref{mobile23}(b) show the successful probability for different $dmax$ and $dmin$. In Fig.\ref{mobile23}(a), we fix $|dmax - dmin| = 5$m and increase $dmin$ from 0m to 5m. The results show that the successful probability decreases rapidly along with $dmin$ which is reasonable due to the larger $dmin$ implying that Willie is far away from Alice.  In Fig.\ref{mobile23}(b), $dmin$ is fixed but $dmax$ grows.  As illustrated in this figure, the successful probability increases with the $dmax$ at first, then decreases. When $dmax\in (2,3)$m ($dmin=0.5$m), the probability reaches its maximal value. Thus, there is a tradeoff between the security level and the value $dmax$ which means that Willie should approach Alice as close as possible and sets his sampling locations in a proper distance, not setting $dmax-dmin$ too great, nor too small. Also, more samples at a location (larger $m$) will result in higher successful probability, but the benefits are trivial compared with other parameters.

\subsection{Informed Jammer}
The previous discussion shows that, Alice cannot hide her transmission behavior in the presence of an active Willie, even if she can utilize other transmitters (or jammers) to increase the interference level of Willie, such as the methods used in \cite{jammer1}\cite{jammer2}
\cite{DBLP:journals/corr/abs-1709-07096}\cite{The_Sound_and_the_Fury}. These methods can only raise the noise level but not change the trend of the sampling values.

The countermeasures discussed above assume that the friendly jammers or interferers in a network are uninformed which means the jammers are not coordinate with Alice. If the jammer has complete knowledge about Alice and Willie, he can closely coordinate with Alice. At the time Alice starts to transmit a codeword, the jammer transmits Gaussian noise simultaneously with the transmit power which is determined according to the distance between him and Alice $d_{a,j}$, the distance between him and Willie $d_{w,j}$, and Alice's transmit power to let Willie's sampling values unchanged when Willie moves along different locations. Willie is then unable to determine that any change has taken place whether Alice is transmitting or silent.

However, as the jammer changes his power to make Willie's sampling value remain the same value whenever Alice is transmitting or not, another eavesdropper can find the change pattern of the jammer's power and then deduce Alice's behavior. Furthermore, this countermeasure is difficult to implement since it is harder to realize the precision synchronizing control between Alice and the jammer.

\section{Willie with Multiple Antennas} \label{ch_multi_ant}
As discussed above, those countermeasures cannot confused Willie effectively. However, it is very difficult for Willie to find Alice's transmission quickly, since Willie has no information about Alice's codebook, he needs to sample at many different locations, especially when Alice's transmission probability is very small. Next we extend our results to a more powerful Willie who has multiple antennas.

\begin{figure} \centering
\centering \epsfig{file=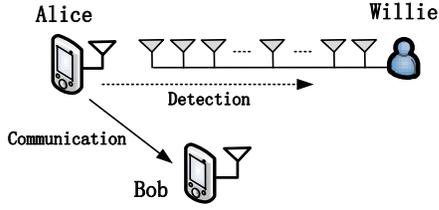, height=1.1in}
\caption{Willie with Multiple Antennas.} \label{multi_ant}
\end{figure}

Different from the framework depicted in Fig. \ref{framework},  Willie arranges multiple antennas at $2t$ different locations to detects Alice's behavior. As shown in Fig.\ref{multi_ant}, those antennas synchronously sample the channel, and Willie uses these samples at different antennas to judge Alice's transmission behavior.

Before diving into details on multiple antennas Willie, the network model and Willie's detection method are in order, as well as discussion of Willie's detection probability.

\subsection{Network Model}
Suppose Alice, Bob, and Willie are placed in a large-scale wireless network, where the locations of transmitters form a stationary Poisson point process (PPP)\cite{Haenggi_PPP} $\Pi=\{X_i\}$ on the plane $\mathbb{R}^2$. The density of the PPP is represented by $\lambda$. Suppose the transmission decisions are made independently across transmitters and independent of their locations for each transmitter, and the transmit power employed for each node are a constant $P_0$. The wireless channel is modeled by large-scale fading with path loss exponent $\alpha$. For simplicity, let the channel gain $\mathbf{h}_{i,j}$ of channel between $i$ and $j$ is static over the signaling period, and all links experience unit mean Rayleigh fading. Then, the aggregated interference seen by node $i$ is the functional of the underlying PPP and the channel gain,
\begin{equation}
  I_i  \equiv  \sum_{k\in\Pi}\sqrt{\frac{P_t}{d_{i,k}^{\alpha}}} \mathbf{h}_{i,k}\cdot s_k\sim \mathcal{N}(0,\sigma^2_{I_i}) \label{eq_2}
\end{equation}
where each $s_k$ is a Gaussian random variable $\mathcal{N}(0,1)$ which represents the signal of the $k$-th transmitter, and
\begin{equation}
\sigma^2_{I_i}=\sum_{k\in\Pi}\frac{P_t}{d_{i,k}^{\alpha}}|\mathbf{h}_{i,k}|^2  \label{eq_4_4}
\end{equation}
are shot noise (SN) process, representing the powers of the interference that Bob and Willie experience, respectively. The Rayleigh fading assumption implies $|\mathbf{h}_{i,j}|^2$ is exponentially distributed with unit mean.

The powers of aggregated interference, $\sigma^2_{I_i}$, is RV which is determined by the randomness of the underlying PPP of transmitters and the fading of wireless channels. The closed-form distribution of the interference is hard to obtain, and its mean is not exist if we employ the unbounded path loss law. We then use a modified path loss law to estimate the mean of $\sigma^2_{I_i}$,
\begin{equation}\label{eq_law}
    l(r)\equiv r^{-\alpha}\mathbf{1}_{r\geq\eta},~~r\in\mathbb{R}_+, ~~\text{for}~\eta\geq 0.
\end{equation}

This law truncates around the origin and thus removes the singularity of impulse response function $l(r)\equiv r^{-\alpha}$. Although transmitters no longer form a PPP under this bounded path loss law (a hard-core point process in this case), this model yields rather accurate results for relatively small guard zones. For $\eta>0$, the mean and variance of $\sigma^2_{I_i}$ are finite and can be given as \cite{Interference_Haenggi}
\begin{eqnarray}
  \mathbf{E}[\sigma^2_{I_i}] &=& \frac{\lambda d c_d}{\alpha-d}\mathbf{E}[\mathbf{|h|^2}]\mathbf{E}[P_t]\eta^{d-\alpha} \\
  \mathbf{Var}[\sigma^2_{I_i}] &=& \frac{\lambda d c_d}{2\alpha-d}\mathbf{E}[{\mathbf{|h|^4}}]\mathbf{E}[P_t^2]\eta^{d-2\alpha}
\end{eqnarray}
where $d$ is the spatial dimension of the network, the relevant values of $c_d$ are: $c_1=2$, $c_2=\pi$, $c_3=4\pi/3$.

When $d=2$, $\alpha=4$, constant transmit power $P_t$, and the fading $|\mathbf{h}|^2\sim\mathrm{Exp}(1)$, we have
\begin{equation}
\mathbf{E}[\sigma^2_{I_i}]=\pi\lambda\eta^{-2}\cdot P_t \label{est}
\end{equation}
which will be used to estimate the interference Willie experiences later.

\subsection{Willie's detection strategy}
Suppose the network is interference limited, i.e., the thermal noise is negligible compared to the aggregated interference from other transmitters. At $i$-th antenna (with the distance $d_i$ between Alice and this antenna) and $(t+i)$-th antenna, Willie's samples are
\begin{eqnarray}
    y_i &=& \sqrt{\frac{P_0}{d_i^{\alpha}}}\cdot s + I_i\sim \mathcal{N}\biggl(0,\frac{P_0}{d_i^{\alpha}}+\sigma^2_{I_i}\biggr) \\
    y_{t+i} &=& \sqrt{\frac{P_0}{d_{t+i}^{\alpha}}}\cdot s + I_{t+i}\sim \mathcal{N}\biggl(0,\frac{P_0}{d_{t+i}^{\alpha}}+\sigma^2_{I_{t+i}}\biggr)
\end{eqnarray}
where $P_0$ is the transmit power of Alice, and $I_i, I_{t+i}$ are the interferences seen by $i$-th antenna and $(t+i)$-th antenna. Even though the interference distribution is identical on the entire plane (in the two-dimensional case), the interference $I_i$ and $I_{t+i}$ are not independent across the plane (which will be discussed later).

With $2t$ sampling values $\mathbf{y}=(y_1^2, y_2^2, \cdots, y_{2t}^2)$ at different antennas, Willie then tries to decide whether $\mathbf{y}$ has a downward trend or not via a Cox-Stuart test. He calculates the differences $\Delta_i=y_i^2-y_{t+i}^2$ ($i = 1..t$), and constructs a test statistic  $\mathbf{T}_{\Delta<0}=\sum^t_{i=1}sgn(\Delta_i)$ (where $sgn(\Delta_i)=1$ for $\Delta_i<0$, $sgn(\Delta_i)=0$ for $\Delta_i\geq 0$). Given a significance level $\beta$ and the binomial distribution $\mathbf{b}\sim b(t,0.5)$, he rejects the null hypothesis $\mathbf{H_0}$ and accept the alternative hypothesis $\mathbf{H_1}$ if $\mathbf{T}_{\Delta<0}<\mathbf{b}(\beta)$.

\subsection{Successful Detection Probability}
The number of negative  differences in $\Delta_1, \Delta_2, ..., \Delta_t$  is $t$ Poisson trials where the success probabilities $\mathbb{P}\{\Delta_i<0\}$ ($i = 1,2,\cdots, t$) differ among the trials, and the number of negative differences in $\Delta_1, \Delta_2, ..., \Delta_t$  can be expressed as follows
\begin{equation}\label{df}
    \mathbf{T}_{\Delta<0}=\sum^t_{i=1}\mathbb{P}\{\Delta_i<0\}=\sum^t_{i=1}\mathbb{P}\{y_i^2<y_{t+i}^2\}
\end{equation}

To calculate $\mathbb{P}\{y_i^2<y_{t+i}^2\}$, we first have to obtain the joint distribution of $(y_i, y_{t+i})$. In the AWGN channels, the samples at $i$-th and $(t+i)$-th antennas, ($y_i$, $y_{t+i}$), have a joint normal distribution, since they are linear combination of the signals from all transmitters in the network, and those signals are independent normal random variables which have the multivariate normal distribution. Therefore the pair ($y_i$, $y_{t+i}$) have the bivariate normal distribution as follows
\begin{equation}\label{normal1}
    (y_i, y_{t+i})\sim\mathcal{N}\biggl(0,0,\frac{P_0}{d_i^{\alpha}}+\sigma^2_{I_i}, \frac{P_0}{d_{t+i}^{\alpha}}+\sigma^2_{I_{t+i}}, \rho_{y_iy_{t+i}} \biggr)
\end{equation}
where $\rho_{y_iy_{t+i}}$ is the correlation coefficient between $y_i$ and $y_{t+i}$, and can be calculated as follows
\begin{eqnarray}
    \rho_{y_iy_{t+i}} & = & \frac{\mathbf{cov}[y_i, y_{t+i}]}{\sqrt{\mathbf{Var}[y_i]}\sqrt{\mathbf{Var}[y_{t+i}]}} \nonumber\\
    & = & \frac{P_0d_i^{-\alpha/2}d_{t+i}^{-\alpha/2}+\rho_{I_iI_{t+i}}\sigma_{I_i}\sigma_{I_{t+i}}}{\sqrt{(P_0d_i^{-\alpha}+\sigma^2_{I_i})(P_0d_{t+i}^{-\alpha}+\sigma^2_{I_{t+i}})}}
\end{eqnarray}
here $\rho_{I_iI_{t+i}}$ is the correlation coefficient between $I_i$ and $I_{t+i}$.

\begin{figure}
\centering \epsfig{file=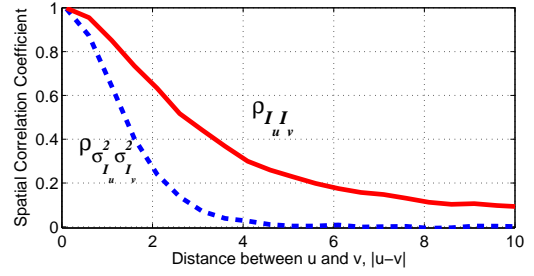, height=1.5in} \caption{The spatial correlations $\rho_{I_iI_{t+i}}$ and $\rho_{\sigma^2_u\sigma^2_v}$ versus the distance between $u$ and $v$. Here the path loss model is $l(x)=\frac{1}{1+\parallel x\parallel^\alpha}$ and the fading $|\mathbf{h}|^2\sim\mathrm{Exp}(1)$.}\label{corr}
\end{figure}

Given any two points $u$ and $v$, using the bounded path loss model $l(x)=\frac{1}{1+\parallel x\parallel^\alpha}$, we can estimate the spatial correlation of $I_u$ and $I_v$, $\rho_{I_uI_v}$, and the spatial correlation of $\sigma^2_{I_u}$ and $\sigma^2_{I_v}$, $\rho_{\sigma^2_{I_u}\sigma^2_{I_v}}$. In Fig.\ref{corr}, the spatial correlations are plotted as a function of the distance between two points $u$ and $v$. Distance decreases the spatial correlation. We observe that,  the decrease of $\rho_{I_uI_v}$ continues at a slower rate than $\rho_{\sigma^2_{I_u}\sigma^2_{I_v}}$, and $\rho_{I_uI_v}>\rho_{\sigma^2_{I_u}\sigma^2_{I_v}}$. This implies that the interference seen by $u$ and $v$ are approximately independent when they are far apart. When they are very close to each other, they experience almost the same interference.

$\rho_{I_iI_{t+i}}$  is hard to obtain. However, the correlation coefficient of $\sigma^2_u$ and $\sigma^2_v$, $u\neq v$ can be calculated as (\cite{Interference_Haenggi}, Lemma 3.3),
\begin{equation}
\rho_{\sigma^2_u\sigma^2_v}=\frac{\int_{\mathbb{R}^2}l(x)l(x-\parallel u-v\parallel)\mathrm{d}x}{\mathbb{E}[{|\mathbf{h}}|^2]\int_{\mathbb{R}^2}l^2(x)\mathrm{d}x}
\end{equation}
where $l(x)$ be the impulse response function to model the path loss attenuation, $\mathbf{h}$ be the fading coefficient. Therefore we can lower bound the correlation between $y_i$ and $y_{t+i}$ as follows
\begin{equation}
    \rho_{y_iy_{t+i}} > \frac{P_0d_i^{-\alpha/2}d_{t+i}^{-\alpha/2}+\rho_{\sigma^2_{I_i}\sigma^2_{I_{i+t}}}\sigma_{I_i}\sigma_{I_{t+i}}}{\sqrt{(P_0d_i^{-\alpha}+\sigma^2_{I_i})(P_0d_{t+i}^{-\alpha}+\sigma^2_{I_{t+i}})}}= \tilde{\rho}_{y_iy_{t+i}}
\end{equation}

Then we have
\begin{eqnarray}
\mathbb{P}\{\Delta_i\leq 0\} & = & \mathbb{P}\{y_i^2\leq y^2_{t+i}\}  \\
                    & = & \iint\limits_{G:x^2<y^2}f_i(x,y)\mathrm{d}x\mathrm{d}y \\
                    & < & \iint\limits_{G:x^2<y^2}\tilde{f}_i(x,y)\mathrm{d}x\mathrm{d}y \label{unequ}
\end{eqnarray}
where $f_i(x,y)$ is the PDF of the bivariate normal variables $(y_i, y_{t+i})$ following distribution of (\ref{normal1}), and $\tilde{f}_i(x,y)$ is the PDF of $(y_i, y_{t+i})$ with different correlation $\tilde{\rho}_{y_iy_{t+i}}$ which is described as follows
\begin{equation}
    (y_i, y_{t+i})\sim\mathcal{N}\biggl(0,0,\frac{P_0}{d_i^{\alpha}}+\sigma^2_{I_i}, \frac{P_0}{d_{t+i}^{\alpha}}+\sigma^2_{I_{t+i}}, \tilde{\rho}_{y_iy_{t+i}} \biggr)
\end{equation}
The unequation in Equ.(\ref{unequ}) is due to $\rho_{y_iy_{t+i}} > \tilde{\rho}_{y_iy_{t+i}}$, and is explained in Appendix B.

Let $\mathbb{P}\{\tilde{\Delta}_i\leq 0\} =\iint\limits_{G:x^2<y^2}\tilde{f}_i(x,y)\mathrm{d}x\mathrm{d}y$, then
\begin{equation}\label{df}
    \mathbf{T}_{\Delta<0}=\sum^t_{i=1}\mathbb{P}\{\Delta_i<0\}<\sum^t_{i=1}\mathbb{P}\{\tilde{\Delta}_i\leq 0\} =\mathbf{T}_{\tilde{\Delta}<0}
\end{equation}
Therefore, we can lower bound the successful detection probability as follows
\begin{eqnarray}
\mathbb{P}\{\mathbf{T}_{\Delta<0}<\mathbf{b}(\beta)\}&>&\mathbb{P}\{\mathbf{T}_{\tilde{\Delta}<0}<\mathbf{b}(\beta)\} \nonumber\\
&=& \mathbb{P}\biggl\{\frac{\mathbf{T}_{\tilde{\Delta}<0}-\mu}{\sigma}   <  \frac{\mathbf{b}(\beta)-\mu}{\sigma }     \biggr\} \nonumber\\
&>& 1-\frac{1}{1+k^2} \label{unequ2}
\end{eqnarray}
where $k=\frac{\mathbf{b}(\beta)-\mu}{\sigma }$, $\mu=\mathbf{E}[\mathbf{T}_{\tilde{\Delta}<0}]=\sum^t_{i=1}\mathbb{P}\{\tilde{\Delta}_i\leq 0\}$, $\sigma^2=\mathbf{Var}[\mathbf{T}_{\tilde{\Delta}<0}]=\sum^t_{i=1}\mathbb{P}\{\tilde{\Delta}_i\leq 0\}\cdot[1-\mathbb{P}\{\tilde{\Delta}_i\leq 0\}]$. Equ. (\ref{unequ2}) is due to Cantelli's Inequality which states that for a random variable $X$ with mean $\mu$ and variance $\sigma^2$, $\mathbb{P}\{X-\mu\geq k\sigma\}\leq 1/(1+k^2)$.

With Equ.(\ref{est}) (with $\eta=1$) as the estimation of the power of interference $\sigma^2_{I_i}$, we can derive $\tilde{\rho}_{y_iy_{t+i}}$, and obtain $\mathbb{P}\{\tilde{\Delta}_i\leq 0\}$, therefore get the lower bound of successful detection probability.

\begin{figure} \centering
\subfigure[]{ \epsfig{file=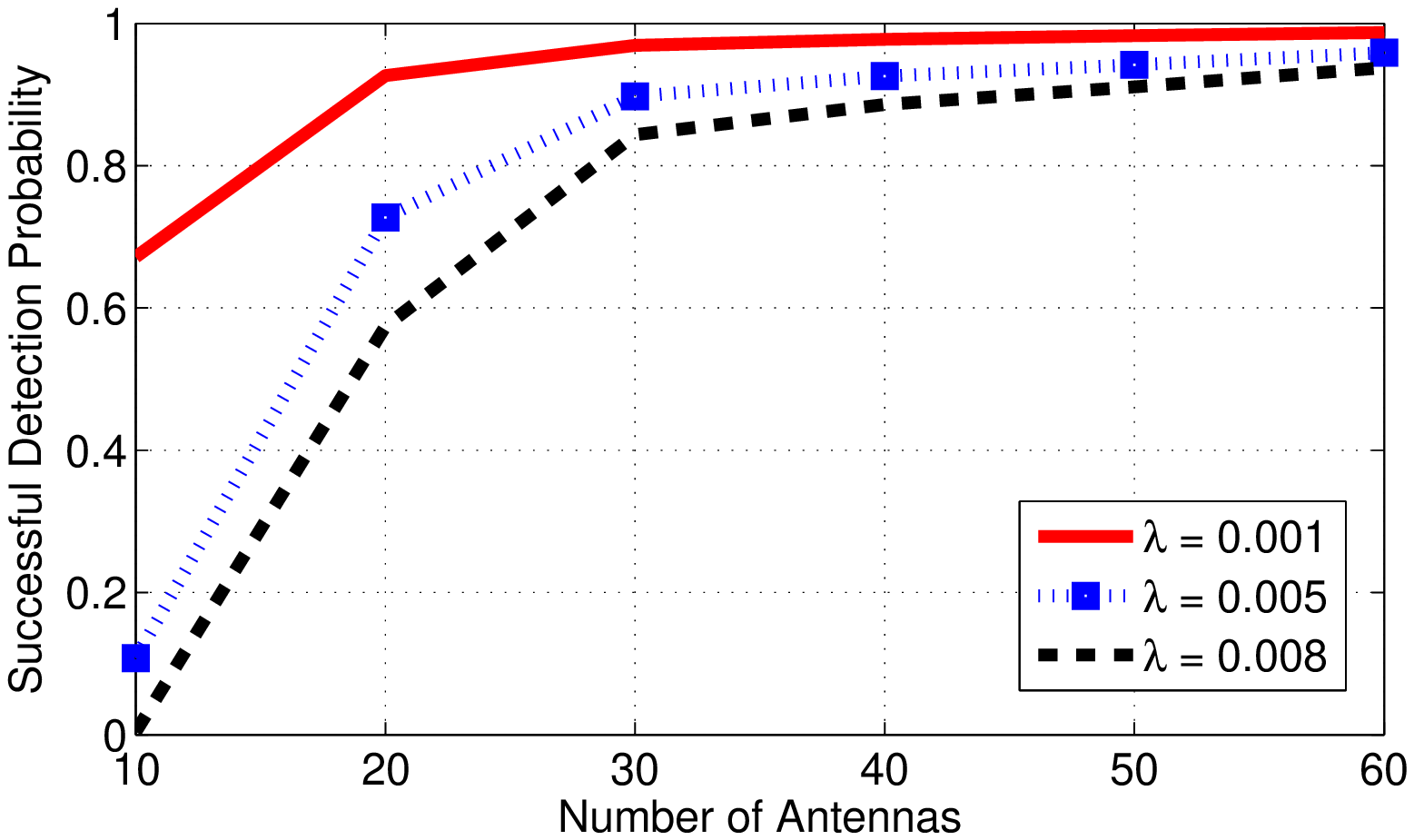, height=1.7in }}
\subfigure[]{ \epsfig{file=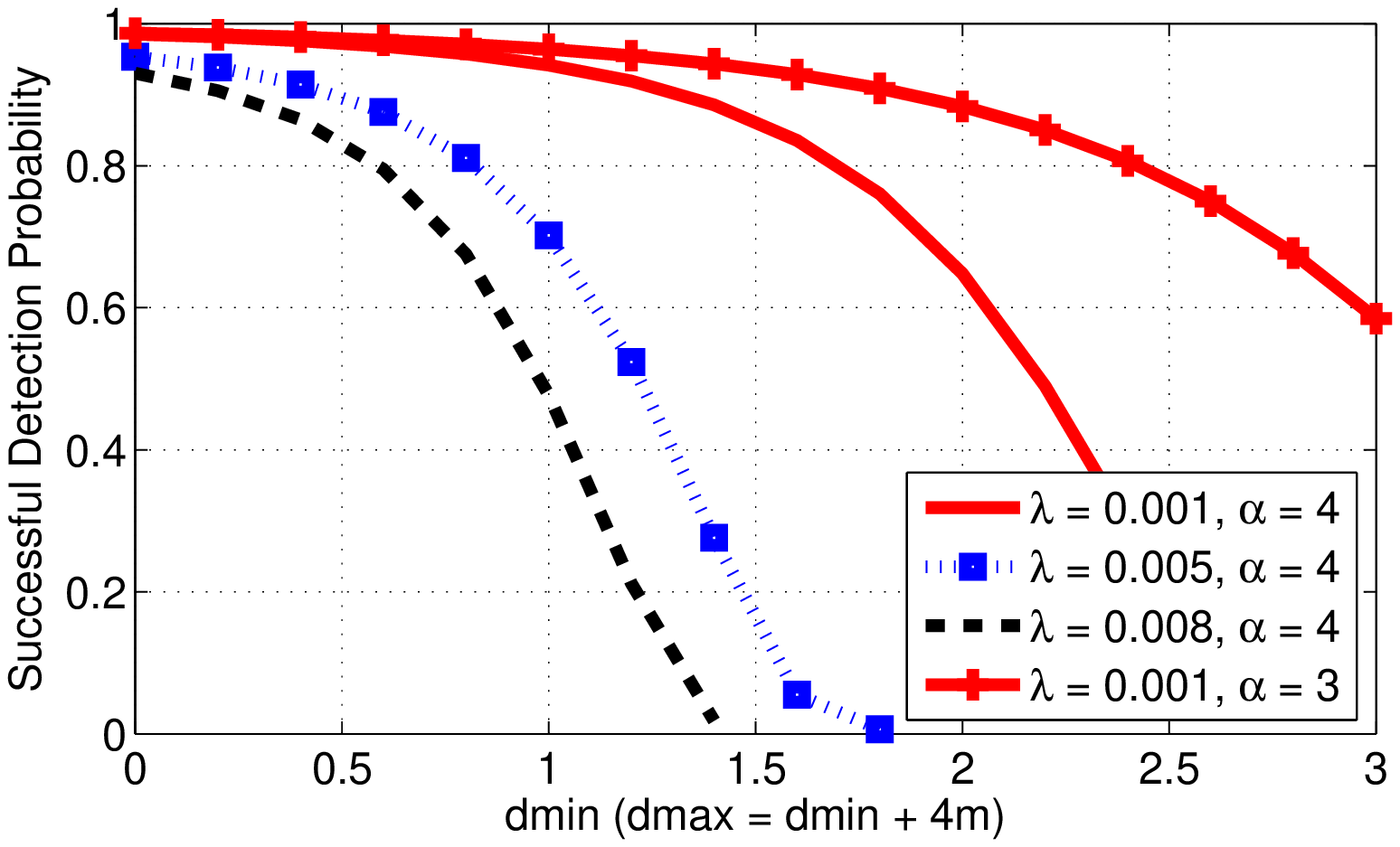, height=1.7in} }
\subfigure[]{ \epsfig{file=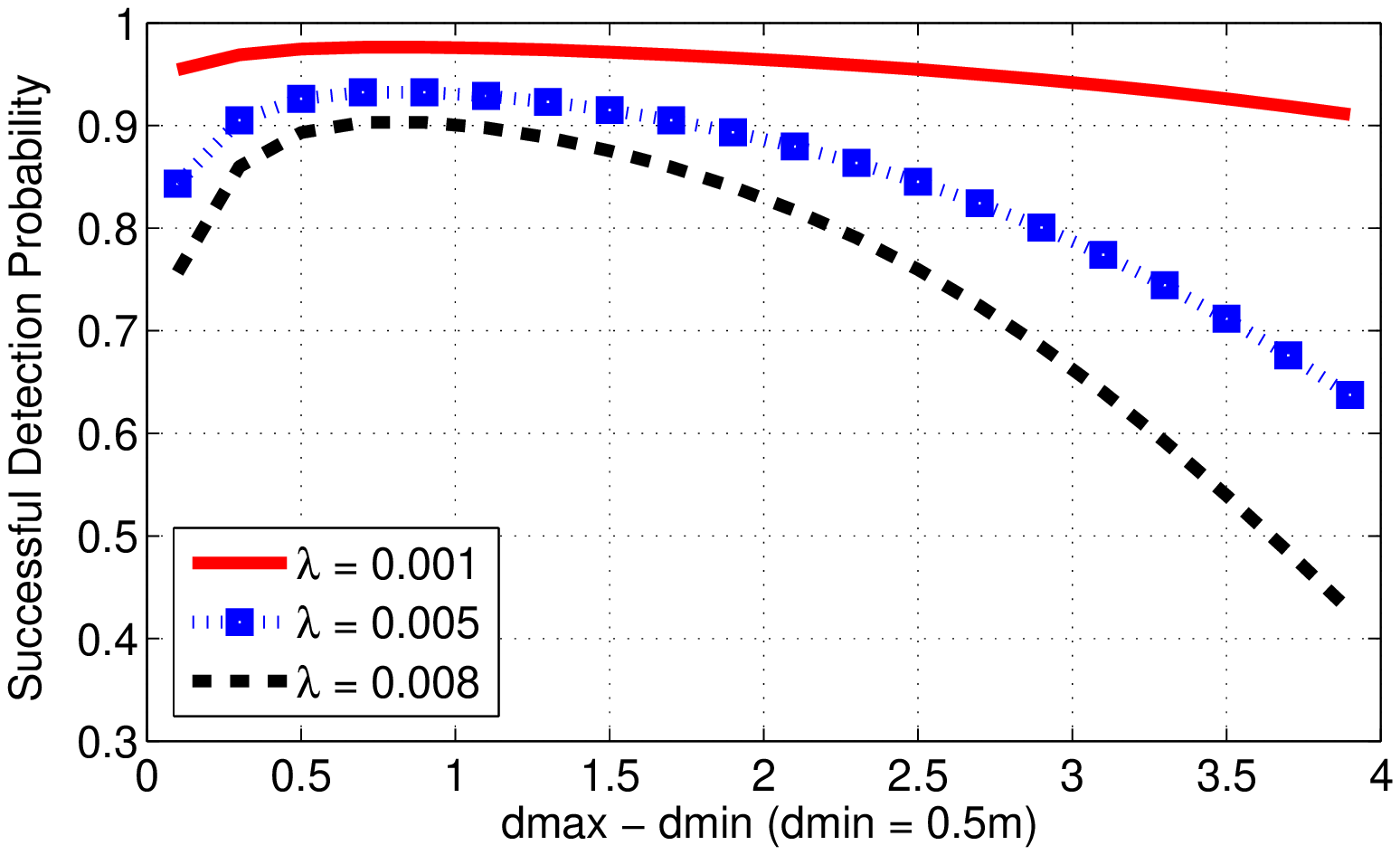, height=1.7in} }
\caption{Willie's successful detection probability with different density $\lambda$ versus (a) the number of antennas with $dmin = 0$m and $dmax = 5$m, (b) $dmin$ with $dmax = dmin + 4$m, (c) $dmax - dmin$ with $dmin$ is fixed to be 0.5m.  Here $\alpha=4$, $\beta=0.05$, and $p_0 = 30$dB.} \label{multi123}
\end{figure}

Fig.\ref{multi123}(a),\ref{multi123}(b), and \ref{multi123}(c) show the successful detection probability with different number of antennas and $dmin$, $dmax$. The graph Fig.\ref{multi123}(a) clearly shows that more antennas will result in higher successful detection probability. Although more dense the network will redult in lower detection probability, Willie can definitely find Alice's transmission at once provided that Willie has enough number of antennas.

In Fig.\ref{multi123}(b), we fix $dmax = dmin + 4$m and increase $dmin$ from 0m to 3m. The results show that the successful probability decreases rapidly along with $dmin$ and more interference from other transmitters (higher $\lambda$) will decrease the the probability. This is reasonable because the larger $dmin$ implies that Willie is far away from Alice. In Fig.\ref{multi123}(c), $dmin$ is fixed but $dmax$ grows. As shown in this figure, the successful probability increases with the $dmax$ at first, then decreases. When $dmax\in (0.5,1)$m ($dmin=0.5$m), the probability reaches its maximal value. Thus, there is a tradeoff between the security level and the value $dmax$ which means that Willie should approach Alice as close as possible and sets his sampling locations in a proper distance.

Therefore, if Willie can deploy multiple antennas and can get close to Alice, he may find Alice's transmission attempt immediately, no need for long sampling. This kind of active Willie is very different to deal with.

\section{Conclusions} \label{ch_6}
We have demonstrated that the active Willie is hard to be defeated to achieve covertness of communications. Alice cannot hide her transmission behavior in the presence of an active
Willie, even if she is placed in a noisy network, or a friendly jammer is involved. A more powerful Willie with multiple antennas can detect Alice's transmission behavior rapidly. Therefore Alice is confronted with enormous challenges if the active Willie is determined to monitor her behavior. As to Alice, there is no better countermeasure to deal with the active Willie in AWSN channels.

As a first step of studying the effects of active Willie on covert wireless communication, this work considers the scenario with one active Willie. A natural future work is to extend the study to multi-Willie. They may work in coordination to enhance their detection ability. Another relative aspect is how to extend the results to DMC or BSC channels, and 5G wireless communication network using beamforming technique and mmWare communication system.


\begin{appendices}
\section{Lindeberg's condition}
Suppose $X_1, X_2,\cdots, X_m$ is a sequence of independent random variables, among them there are $mr$ $(0\leq r\leq 1)$ random variables obeying distribution $\sigma^2_W\cdot \chi^2(1)$, the remain are $\sigma^2_0\cdot \chi^2(1)$ random variables, where $\sigma^2_W$ and $\sigma^2_0$ are finite value. Define
$$s^2_m=\sum_{i=1}^m \mathbf{Var}[X_i]=mr\cdot 2\sigma_W^4 + m(1-r)\cdot 2\sigma_0^4$$

According to Chebyshev's Inequality,
$$\mathbf{P}\{|X_i-\mu_i|>\epsilon s_m\}<\frac{\mathbf{Var}[X_i]}{\epsilon^2 s_m^2}$$
we have
\begin{eqnarray*}
  f(m) &=& \frac{1}{s^2_m}\sum_{i=1}^m \mathbf{E}[(X_i-\mu_i)^2\cdot \mathbf{1}_{\{|X_i-\mu_i|>\epsilon s_m\}}] \\
   &<& \frac{1}{s^2_m}\biggl[mr\cdot \frac{4\sigma_W^8}{\epsilon^2 s_m^2} + m(1-r)\cdot \frac{4\sigma_0^8}{\epsilon^2 s_m^2} \biggr] \\
   &=& \frac{r\cdot\sigma_W^8 + (1-r)\cdot\sigma_0^8}{m\epsilon^2[r\cdot\sigma_W^4 + (1-r)\cdot\sigma_0^4]^2}
\end{eqnarray*}

Therefore for every $\epsilon>0$,
$$\lim_{m\rightarrow\infty}\frac{1}{s^2_m}\sum_{i=1}^m \mathbf{E}[(X_i-\mu_i)^2\cdot \mathbf{1}_{\{|X_i-\mu_i|>\epsilon s_m\}}] = 0$$
which means that Lindeberg's condition holds, i.e., the distribution of
$$\frac{\sum_{i=1}^m X_i - \sum_{i=1}^m \mathbf{E}(X_i)}{\sqrt{\sum_{i=1}^m \mathbf{Var}(X_i)}}$$
converges towards the standard normal distribution $\mathcal{N}(0,1)$.

\section{The probability $\mathbb{P}\{X^2<Y^2\}$ and correlation coefficient $\rho$ for a bivariate normal distribution}
For a bivariate normal distribution with $\mu_1=\mu_2=0$,
\begin{equation*}
    (X,Y)\sim\mathcal{N}(0,0,\sigma_1^2,\sigma_2^2,\rho)
\end{equation*}
if $\sigma_1^2>\sigma_2^2$, then the probability $\mathbb{P}\{X^2<Y^2\}$ decreases with $|\rho|$; if $\sigma_1^2<\sigma_2^2$, $\mathbb{P}\{X^2<Y^2\}$ increases with $|\rho|$. In the case $\sigma_1^2=\sigma_2^2$, $\mathbb{P}\{X^2<Y^2\}=1/2$.

Because the analytical expressions of $\mathbb{P}\{X^2<Y^2\}$ is hard to obtain, we can verify the above results through numerical results. The probability $\mathbb{P}\{X^2<Y^2\}$ can be calculated as follows
\begin{equation*}
    \mathbb{P}\{X^2<Y^2\} = \iint\limits_{G:x^2<y^2}f(x,y)\mathrm{d}x\mathrm{d}y
\end{equation*}
where $f(x,y)$ is the PDF of $(X,Y)$,
\begin{eqnarray*}
    f(x,y)&=&\frac{1}{2\pi\sigma_1\sigma_2\sqrt{1-\rho^2}} \\
          & & \times\exp{\biggl\{\frac{-1}{2(1-\rho^2)}\biggl[\frac{x^2}{\sigma_1^2}-2\rho\frac{x\cdot y}{\sigma_1\sigma_2}+\frac{y^2}{\sigma_2^2}\biggr]\biggr\}}
\end{eqnarray*}

\begin{figure} \centering
\subfigure[]{ \epsfig{file=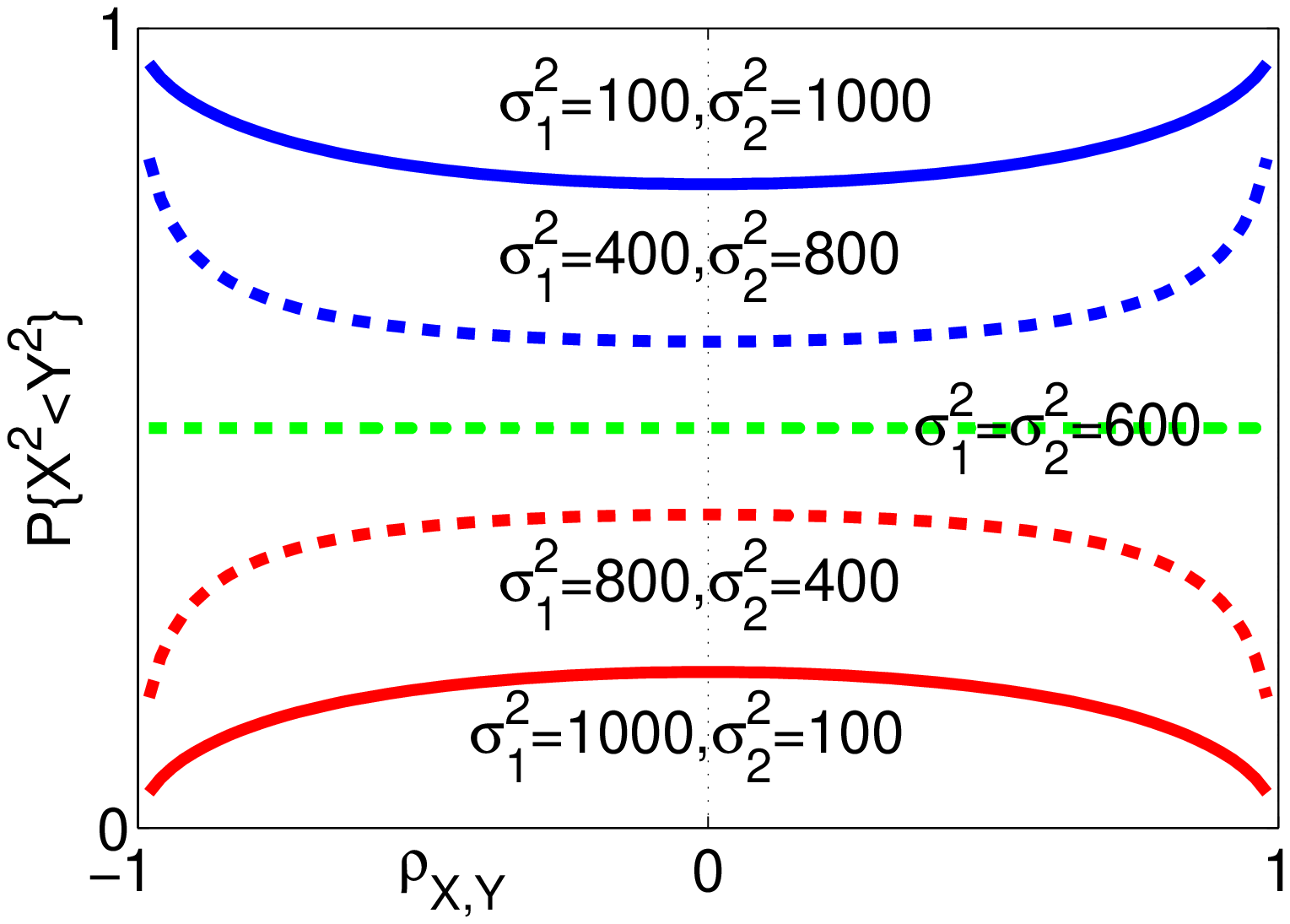, height=0.85in }}
\subfigure[]{ \epsfig{file=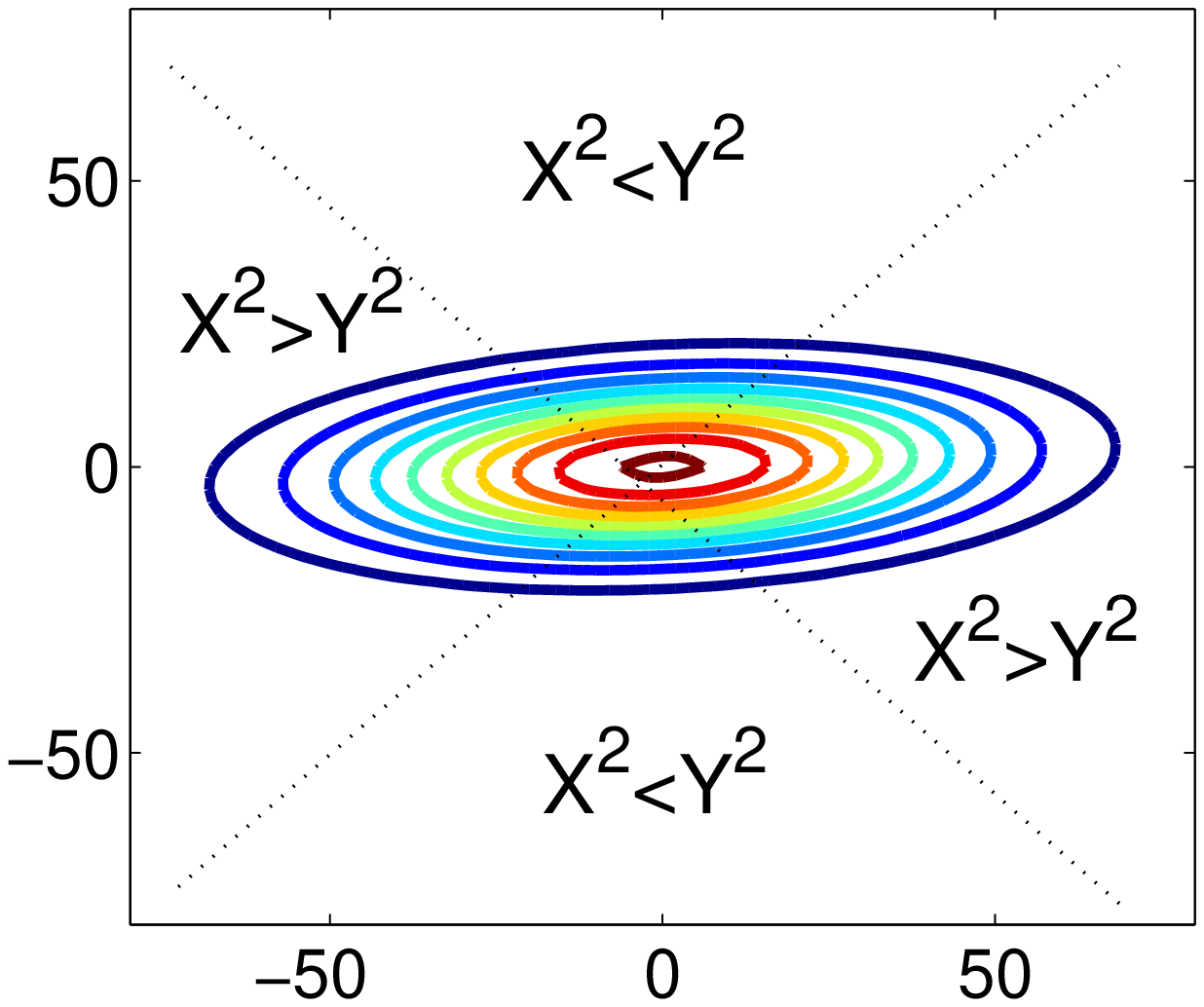, height=0.8in} }
\subfigure[]{ \epsfig{file=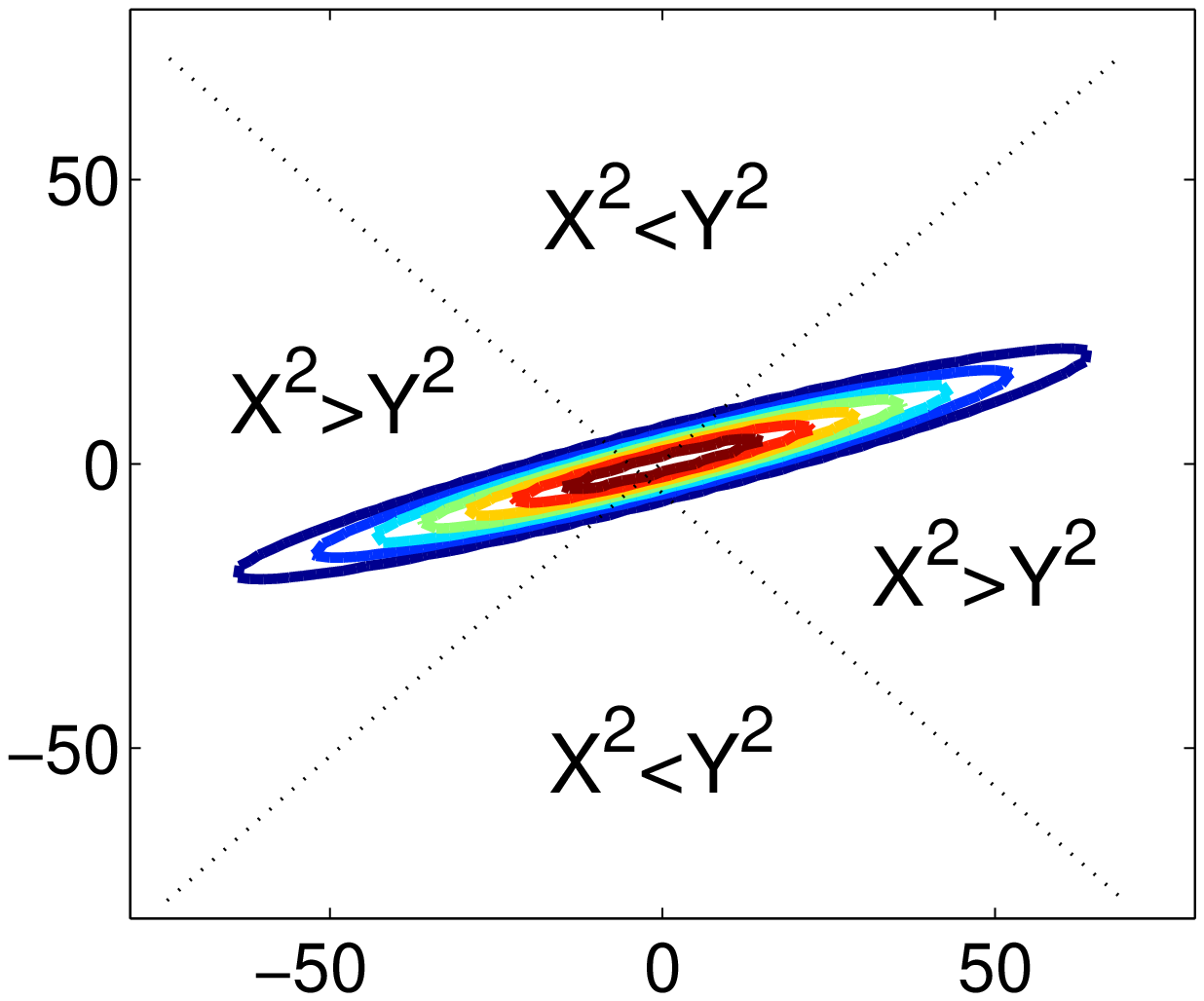, height=0.8in} }
\caption{(a) The probability $\mathbb{P}\{X^2<Y^2\}$ and correlation coefficient $\rho_{X,Y}$ for a bivariate normal distribution. (b) The contour of PDF of $(X,Y)\sim\mathcal{N}(0.0,1000,100,0.1)$. (c) The contour of PDF of $(X,Y)\sim\mathcal{N}(0.0,1000,100,0.9)$.}
\label{rho}
\end{figure}

As depicted in Fig.\ref{rho}(b)(c), when $\sigma_1^2>\sigma_2^2$, if $\rho$ increases, the distribution of $(X,Y)$ is more focused on the region $G: X^2>Y^2$, resulting in less $\mathbb{P}\{X^2<Y^2\}$. Fig.\ref{rho}(a) also illustrates this result, e.g., $\mathbb{P}\{X^2<Y^2\}$ decreases with $|\rho|$ when $\sigma_1^2>\sigma_2^2$, and increases with $|\rho|$ when $\sigma_1^2<\sigma_2^2$.

\end{appendices}


\end{document}